\documentclass[USenglish,oneside,twocolumn]{article}
\usepackage{caption}

\usepackage[linesnumbered,noline,ruled]{algorithm2e}
\usepackage{paralist}
\usepackage{citesort}
\usepackage{url}
\usepackage{amssymb,amsmath,amsthm, amsfonts}
\usepackage{mathtools}
\usepackage{color}
\usepackage{authblk}
\usepackage{subfigure}
\usepackage{natbib}
\usepackage{enumitem}
\usepackage{tikz}
\usepackage{pgfplots}

\setlist[itemize]{noitemsep, nolistsep}
\setlist[enumerate]{noitemsep, nolistsep}
\newtheorem{theorem}{Theorem}
\newtheorem*{thm}{Theorem}

\newenvironment{proofsketch}{\par\noindent\textsl{Proof (sketch):}\kern1ex}%
{\hfil\hskip2em\penalty250\parfillskip=0pt\finalhyphendemerits=0$\qed$%
	\par\smallskip\par}

\newtheorem*{obs}{Observation}
\newtheorem*{definition}{Definition}
	
\newenvironment{myquote}%
  {\smallskip \list{}{\leftmargin=0.2in\rightmargin=0.2in}\item[] \em}%
  {\endlist}

\def\O{\mathbb{O}}

\usepackage{etoolbox}
\patchcmd{\paragraph}{\@parfont}{\bfseries}{}{} 
\patchcmd{\paragraph}{\parindent}{0pt}{}{} 
\renewcommand{\paragraph}[1]{\medskip \noindent {\bf #1.}~}

\newcommand{\BT}{\begin{theorem}}
\newcommand{\ET}{\end{theorem}}
\newcommand{\BD}{\begin{definition}}
\newcommand{\ED}{\end{definition}}
\newcommand{\BCR}{\begin{corollary}}
\newcommand{\ECR}{\end{corollary}}
\newcommand{\BEX}{\begin{example}}
\newcommand{\EEX}{\end{example}}
\newcommand{\BL}{\begin{lemma}}
\newcommand{\EL}{\end{lemma}}
\newcommand{\BP}{\begin{proposition}}
\newcommand{\EP}{\end{proposition}}
\newcommand{\BCM}{\begin{claim}}
\newcommand{\ECM}{\end{claim}}
\newcommand{\BPF}{\begin{proof}}
\newcommand{\EPF}{\end{proof}}
\newcommand{\BEN}{\begin{enumerate}}
\newcommand{\EEN}{\end{enumerate}}

\newcommand{\BI}{\begin{itemize}}
\newcommand{\EI}{\end{itemize}}

\newcommand{\BO}{\begin{observation}}
\newcommand{\EO}{\end{observation}}

\newcommand{\BDS}{\begin{description}}
\newcommand{\EDS}{\end{description}}

\newcommand{\floor}[1]{{\left\lfloor{#1}\right\rfloor}}

\renewcommand{\>}{{\rangle}}

\def\squareforqed{\hbox{\(\blacksquare\)}}
\def\qed{\hfill \squareforqed}

\newcommand{\secparam}{\ensuremath{\lambda}}

\newcommand{\ignore}[1]{}

\newcounter{defcounter}
\newlength{\protowidth}

\newcommand{\olrk}[1]{%
   \ifx\nursymbol#1\else\!\!\mskip4.5mu plus 0.5mu\left(#1\right)\fi}
\newcommand{\elrk}[1]{%
   \ifx\nursymbol#1\else%
        \!\!\mskip4.5mu plus0.5mu\left[\mskip2.5mu plus0.5mu #1\right]\fi}
\def\path{\ensuremath{{\sf location}}}

\def\Access{\mathsf{PhysicalAcc}}

\def\Access{\ensuremath{{\sf Access}}}

\newcommand{\sys}{SqORAM}
\newcommand{\usnaoram}{DetWoORAM}

\begin{document}
\title{\bf SqORAM: Read-Optimized Sequential Write-Only Oblivious RAM}
%\runningtitle{SqORAM}
\author[1]{Anrin Chakraborti}
\author[1]{Radu Sion}
\affil[1]{Department of Computer Science, Stony Brook University}
\date{}
\maketitle

%\runningtitle{SqORAM}

\abstract{
	Oblivious RAM protocols (ORAMs) allow a client to access data from an untrusted storage device without revealing the access patterns. Typically, the ORAM adversary can observe both read and write accesses. Write-only ORAMs target a more practical, {\em multi-snapshot adversary} only monitoring client {\em writes} -- typical for plausible deniability and censorship-resilient systems~\cite{hive,datalair}.
	
This allows write-only ORAMs to achieve significantly-better asymptotic performance. However, these apparent gains do not materialize in real deployments primarily due to the random data placement strategies used to break correlations between logical and physical namespaces, a required property for write access privacy.  Random access performs poorly on both rotational disks and SSDs (often increasing wear significantly, and interfering with wear-leveling mechanisms).

In this work, we introduce {\sys}, a new locality-preserving write-only ORAM that preserves write access privacy without requiring random data access. Data blocks close to each other in the logical domain land in close proximity on the physical media.  Importantly, {\sys} maintains this data  locality property over time, significantly increasing read throughput.

A full Linux kernel-level implementation of {\sys} is 100x faster than non locality-preserving solutions for standard workloads and is 60-100\% faster than the state-of-the-art for typical file system workloads.
% and is only 1.5x slower than a baseline block device implementation with no access privacy.
}

\section{Introduction}
\label{seqoram:intro}

Dramatic advances in storage technology have resulted in users storing personal (sensitive) information on portable devices (mobiles, laptops etc). 
% Unfortunately, the stored information becomes vulnerable to unauthorized disclosures through attacks on the storage medium. 
To ensure confidentiality, data can be encrypted at-rest.  However, often this is not enough since access sequences of read and written locations leaks significant amounts of information about the data itself, defeating the protection provided by encryption \cite{accesspatternleak}. 

%\paragraph{Oblivious RAM (ORAM)}
% 
To mitigate this, one solution is to access items using an oblivious RAM (ORAM) protocol to hide data access patterns from an adversary monitoring the storage device or unsecured RAM. Informally, ORAM protocols ensure (computational) indistinguishability between multiple equal-length query/access sequences. 
% for accessing 
%items from an untrusted storage media, while effectively hiding  

% {\color{blue}
ORAMs have been typically proposed in the context of securing untrusted remote storage servers e.g., public clouds. However, protecting data access patterns is a key requirement in many other privacy-critical applications \cite{hive,datalair,usna_stash_free_oram} including plausibly-deniable storage systems. Plausible deniability ultimately aims to enable users to deny the very existence of sensitive information on storage media when confronted by coercive adversaries e.g., border officers in oppressive regimes. This is essential in the fight against increasing censorship and intrusion into personal privacy \cite{defy,chenPets19}

Unfortunately, it is impractical to deploy existing ORAM mechanisms in such systems due to prohibitively-high access latencies deriving from high asymptotic overheads for accessing items and ORAM-inherent randomized access patterns.

We note however that for plausible deniability, a full ORAM protocol protecting access patterns of all operations in real time may be unnecessary. After all, most realistic adversaries in this context (e.g., border officers, hotel maids etc) only have {\em multi-snapshot} capabilities and can only access the device periodically while at rest. As a result, except in the presence of end-point compromise (e.g., malware) in which case the adversary usually gets full access to the entire system, in a typical scenario, runtime reads and {\em writes} are not observed \cite{hive}. 

This opens up a significant opportunity, namely the idea of a write-only ORAM, a simpler, more effective ORAM that only protects the privacy of write accesses \cite{li_write_only}. In fact write-access privacy has been proven to be one of the necessary and sufficient requirements for building strong plausible deniability system\cite{datalair}. Write-only ORAMs have been employed to build state-of-the-art plausibly-deniable storage mechanisms including systems such as HIVE \cite{hive} and DataLair \cite{datalair}. 

Write-only ORAMs can achieve significantly better asymptotic performance compared to full ORAAMs. Yet, existing designs are still orders of magnitude slower than the underlying raw media. For example, HIVE is almost {\em four orders of magnitude slower} than HDDs and {\em two orders of magnitude slower} than SSDs.
%}

The main contributor to this slowdown is clear: random placement of data meant to break {\em linkability} between separate writes \cite{usna_stash_free_oram,chenPets19}, an important property ensuring that an adversary cannot link a set of writes to each other logically, given multiple snapshots of the media. Random data placement results in dramatically increased disk-seek related latencies. Non-locality of access also interferes with numerous higher-level optimizations including caching policies, read-aheads etc.

To mitigate this, \citet{usna_stash_free_oram} recently proposed a write-only ORAM that preserves locality of access for writes. This is based on the idea that {\em unlinkability} can also be achieved by writing data to the storage media in a canonical form -- e.g., by writing logical data blocks sequentially at increasing physical block addresses, independent of their logical block addresses, similar to a log-structured file system.

Yet, this unfortunately does not solve the problem. Sequentially-written physical blocks rarely translate in locality for the logically-related items. In fact, sequential-write log structured file systems perform quite poorly for {\em reads} as logically related data ends up scattered across the disk over time \cite{betrfs}. 

And since reads tend to dominate in modern workloads -- e.g., over 70\% of ext4 requests are reads  \cite{survey_network_FS,survey_typical_FS} -- optimizing for logical {\em reads} is especially important. 

%Even web-hosted services such as mail servers, follow similar read-write distributions \cite{web_server_read_write_dist}. 

This paper introduces the philosophy, design and implementation of {\sys}, a new write-only ORAM system that preserves locality of access for {\em both reads and writes.} {\sys} introduces a seek-optimized data layout: logically-related data is {\em initially placed} and then {\em maintained throughout its lifetime} in close proximity on the underlying media, to dramatically increase locality of access and overall throughput, especially when paired with standard file systems.

\paragraph{Locality-Preserving Hierarchical Layout}
{\sys} smartly adapts hierarchical ORAM \cite{goldreich} techniques for periodic efficient reshuffles keeping
logically-related items in close proximity. 

Specifically, hierarchical ORAMs store blocks in multiple {\em levels}, where each level is twice the size of the previous level. Blocks are stored at random locations in a level and the contents in each level are periodically reshuffled and moved to the next level. In {\sys} data is organized similar to hierarchical ORAMs. However, instead of randomized block placement, {\sys} \emph{stores blocks per-level sorted on their logical address}. Related blocks with logically contiguous addresses are stored close together in the levels, and can be fetched efficiently while eliminating seek-related latencies.

A key set of insights underlies this: (i) in standard ORAMs, randomized block placement is mainly necessary to protect the privacy of read patterns, and (ii) storing blocks in a standard canonical form (e.g., sorted on logical address) does not leak write access pattern privacy. 

\paragraph{Asymptotically-Efficient Level Reshuffles}
In standard hierarchical ORAMs, level reshuffles are expensive. Random block placement (and read-privacy guarantees) necessitates complex oblivious sorting based mechanisms to securely reshuffle data. Eliminating read-privacy requirements provides the opportunity for simple and asymptotically more efficient (by a factor $\O(\log N)$) level reshuffles. 
%{\sys} employs an efficient, seek-optimized level reshuffle mechanism.

\paragraph{Efficiently Tracking Blocks}
In hierarchical ORAMs locating a particular block on disk is expensive and requires searching in {\em all} the levels. This is necessary because the location of a block (the level that it is read from) reveals information about its last access time, thus all levels need to be searched to prevent the server from learning the one of interest.

{\sys} does not face this requirement and employs a new efficient mechanism to securely and efficiently track the location of blocks based on last access times. To retrieve a block, only one level needs to be searched -- this includes an index lookup and reading the block from its current on-disk location.

% \todo{retrieved or identified? are you talking about the lookup ? we still need another access after this or is the block there?} efficiently with a single level lookup avoiding searches in all other levels. 
% }

% Importantly, the frequency of reshuffles and the corresponding access patterns does not leak write-access privacy.

%
%This preserves locality of accesses for writes without sacrificing privacy. 
%
%
%
%Yet, sequentially writing data alone is not enough to guarantee that logically-related items remain close together on disk -- sequential-write log structured file systems perform poorly for {\em reads} as logically related data ends up scattered across the disk over time \cite{betrfs}.  
%
%To achieve significantly higher overall throughput, locality for {\em reads} is especially important, since many workloads predominantly include reads. In fact, typically, for standard file systems (e.g., ext4), over 70\% of requests are typically reads and 30\% are writes \cite{survey_network_FS,survey_typical_FS}. Even web-hosted services such as mail servers, follow similar read-write distributions \cite{web_server_read_write_dist}. 

%\subsection{Motivation \& Insights}

%
%\paragraph{Locality for Reads}
%

%
% We address the issue of read locality for write-only ORAMs in {\sys}. 

%As a result of this data-placement strategy, when compared with HIVE-ORAM \cite{hive}, a
%publicly available write-only ORAM implemented for the Linux
%Kernel~\cite{hive}, that does not preserve locality of accesses, {\sys} read
%and write throughputs are orders of magnitude faster for sequential
%accesses.
\paragraph{Evaluation}
Compared to randomization-based write-only ORAMs (HIVE \cite{hive}, DataLair \cite{datalair}) that have been employed in plausible-deniability schemes, {\sys} is orders of magnitude faster for both sequential reads and writes.  Compared to the state-of-the-art \cite{usna_stash_free_oram}, 
{\sys} features a 2x speedup for sequential reads and achieves near raw-disk throughputs in the presence of
extra memory. As an application, 
experiments demonstrate that {\sys} is faster than \cite{usna_stash_free_oram} for a typical file system workload with 70\% reads and 30\% writes~\cite{survey_network_FS,survey_typical_FS} and is only 1.5x slower than the baseline.

\section{Related Work}
\label{seqoram:related}
\vspace{-5pt}

\citet{li_write_only} proposed the first write-only ORAM
scheme with an amortized write complexity of $\O(B \times \log{N})$
where $B$ is the block size of the ORAM and $N$ is the number of blocks in
the ORAM.  
%Read complexity is $\O(N)$ without $\O(N)$
%local storage.

\citet{hive} designed a constant time write-only ORAM
scheme assuming an $\O(\log N)$ sized stash stored in memory (HIVE-ORAM). 
It maps data from a logical address space uniformly randomly to the physical
blocks on the underlying device.  \citet{datalair} improved upon the HIVE-ORAM construction by reducing the overall access complexity by a factor of $\O(\log N)$.  The ORAM 
utilizes multiple oblivious data structures to optimize the expensive HIVE-ORAM free block finding process from \cite{hive}.

%The position map containing this mapping
%is then recursively stored in $\O(\log N)$ smaller ORAMs, a standard
%technique introduced in \cite{Shi_obliviousram}.  The recursive technique
%reduces the logical block access complexity for the position map by storing
%the position map blocks in logical blocks of {\em smaller} sizes.  Under
%this assumption, HIVE-ORAM~\cite{hive} accesses a {\em constant} number of
%logical blocks at the cost of some overflow that is
%stored in the in-memory stash.

% {\color{blue}
%\paragraph{Locality-preserving Write-only ORAM}
%
\citet{usna_stash_free_oram} recently proposed DetWoORAM, a write-only ORAM that is optimized for sequential writes with $\O(\log{N})$ read complexity and $\O(1)$ write complexity. 
The idea is to write blocks to the disk sequentially, at increasing physical addresses, independent of their logical address, not unlike log-structured file systems. It has been shown \cite{usna_stash_free_oram} that maintaining this layout ensures that a multi-snapshot adversary cannot link a set of writes to each other logically given multiple snapshots of the disk.
%
%
%
% A particular advantage of this approach is that if the accesses in the logical domain are also mostly sequential or have high degree of locality, e.g., due to a standard overlying file system, then the physical layout maintains some of that locality to the benefit of subsequent reads. 
However, once written to disk, blocks are not guaranteed to remain at the same location and, on updates, blocks are written to new locations (e.g., at the head of the log), thus destroying locality of logical accesses for subsequent reads. 
% }

%The idea is to organize the layout of the ORAM as an append-only log and performing the {\em next} write as an append to the log. Although, this significantly improves write throughput, logically related data is not maintained together on the disk (unlike standard file systems) and thus not read-optimized.

%In effect, DetWoORAM reaps the benefits of fast writes by structuring data on the disk similar to log-structured file system. 

%Write-only ORAMs have been previously used to build oblivious data backup mechanisms \cite{oblivisync}  and secure processors \cite{flatoram}.
\citet{oblivisync} leveraged write-only ORAMs to build an oblivious file system to backup data on a remote storage server while \citet{flatoram} proposed a secure processor design leveraging write-only ORAMs. 

%Since these goals are not directly related to our motivation, we refer to the respective papers for further details.

\section{Background}
\label{sqoram:background}

\paragraph{Adversary}
%
% {\color{blue} 
Realistic adversaries have multi-snapshot capabilities. They can observe the storage media not just once but at multiple different times and possibly take snapshots, maybe after {\em every} write operation. 
They may save these snapshots -- including device-specific information, filesystem metadata and bits stored in each block -- and compare past snapshots with the current state in an attempt to learn about the location of the written information.  
% }

\paragraph{Security Definition}
To hide access patterns from a multi-snapshot adversary, a write-only ORAM needs to ensure write-access privacy.

\begin{definition}[Write-Access Privacy]
	\label{def:privacy}
	Let $\vec{y} = (y_1, y_2, \ldots)$ denote a sequence of
	operations, where each $y_i$ is a ${\sf Write}(a_i, d_i)$; here, $a_i \in [0, N)$ denotes the logical address of the block being written, and $d_i$ denotes a block of data being written. For an ORAM scheme $\Pi$, let $\Access^\Pi(\vec{y})$ denote the
	physical access pattern that its access protocol produces for the logical
	access sequence $\vec{y}$.  We say the scheme $\Pi$ ensures {\em write-access privacy} if for any
	two sequences of operations $\vec{x}$ and $\vec{y}$ of the same length, it
	holds
	$$ \Access^\Pi(\vec{x}) ~~ \approx_c ~~ \Access^\Pi(\vec{y}), $$
	where $\approx_c$ denotes computational indistinguishability (with respect to
	the security parameter $\secparam$).
\end{definition}

%\paragraph{Locality \& Security}
%%
%The most important requirement of a write-only ORAM scheme is to ensure write-unlinkability which ensures that a multi-snapshot adversary cannot link a set of writes to each other logically, given multiple snapshots of the disk. This is achieved by placing data block on-disk at physical locations independent of its logical address. 
%
%Recently, \citet{usna_stash_free_oram} showed that write-unlinkability can be achieved by writing every block
%to {\em sequentially} increasing physical addresses e.g., to the head of an append-only log in case of a log-structured file system. This does not sacrifice write-access privacy and also provides opportunities for performing locality-preserving writes \cite{usna_stash_free_oram}.

%One of the ways to accomplish this is by writing blocks to random locations \cite{hive,datalair}, which performs poorly on hardware due to random placement of data. 

\subsection{Hierarchical ORAM}
\label{sqoram:background:hierarchical_oram}

In this section, we review hierarchical ORAM constructions, which is an important building block of {\sys}.

\paragraph{Organization}
Hierarchical ORAMs \cite{goldreich} organize data into \emph{levels}, with each level twice the size of the previous level. Specifically, for a database with $N$ data blocks, the ORAM consists of $\log N$ levels, with level $i \leq \log N$ containing 
$2^{i}$ blocks (including dummy blocks).

At each level, blocks are stored at uniformly random \emph{physical} locations determined by applying level-specific hash functions on logical addresses. In other words, each level storage can be viewed as a hash table of appropriate size \cite{kushilevitzoram,bforam}. Blocks are always written to the top level first and periodically move down the levels as a result of reshuffles. 

% The first hierarchical construction was provided by
%Goldreich and Ostrovsky  and achieves an amortized
%communication complexity of $\O(\log^{3}N)$ data blocks per access and
%requires only logarithmic storage at the client side.

\paragraph{Queries}
During queries, all levels are searched sequentially for the target block using the level-specific hash functions to determine the exact location of the block in a particular level. When a block is found at a certain level $i$, \emph{dummy} blocks are read from rest of the levels.

After a block is read from a certain level in the ORAM, it is written re-encrypted to the top level. Once the top level is full, the contents of the top level are securely reshuffled and written to the next level. This mechanism is applied for 
all levels in the ORAM.

\paragraph{Reshuffles}
The most expensive step of hierarchical ORAMs is the level reshuffle. This is because when reshuffling level $i$, its contents are obliviously sorted on randomly-assigned tags (e.g., logical address hashes) and written to random locations in level $i+1$. Consequently, reshuffles are expensive not only in terms of the total I/O, but also in the number of seeks.

%Various mechanisms have been proposed to make the reshuffle more efficient. 
%Williams and Sion~\cite{usablepir} show how to achieve an amortized
%construction with $\O(log^{2}N)$ communication complexity with
%$\O(\sqrt{N})$ client storage using an oblivious merge sort.  Pinkas
%{\em et al.} \cite{Pinkasoram} use {\em cuckoo hashing} and randomized shell
%sort \cite{randomizedshellsort} over the original Goldreich and Ostrovsky
%solution \cite{goldreich} and achieve an amortized communication complexity
%of $\O(log^{2}N)$ with constant client side storage.  Goodrich {\em et
%al.} \cite{goodrich_deamortized} show how to de-amortize the original square
%root solution and hierarchical solution \cite{goldreich} and achieve a
%worst-case complexity of $\O(logN)$ in the presence of $\O(n^{r})$
%client side storage where $r > 0$.  In \cite{kushilevitzoram}, Kushilevitz
%{\em et al.} use cuckoo hashing and rotating buffers to provide a
%de-amortized construction of the original hierarchical solution
%\cite{goldreich} which achieves a worst-case communication complexity of
%$\O(\frac{\log^{2}N}{\log\log N})$.  Unlike the de-amortization techniques used in
%\cite{kushilevitzoram,goodrich_deamortized} where each query performs an
%additional fixed amount of work for the reshuffle, PD-ORAM \cite{privatefs}
%provides a way to de-amortize the level construction with parallel clients
%that perform one (or more) level reshuffles in the background.

\section{Overview}

\begin{figure*}[h!]
	\includegraphics[scale=0.13]{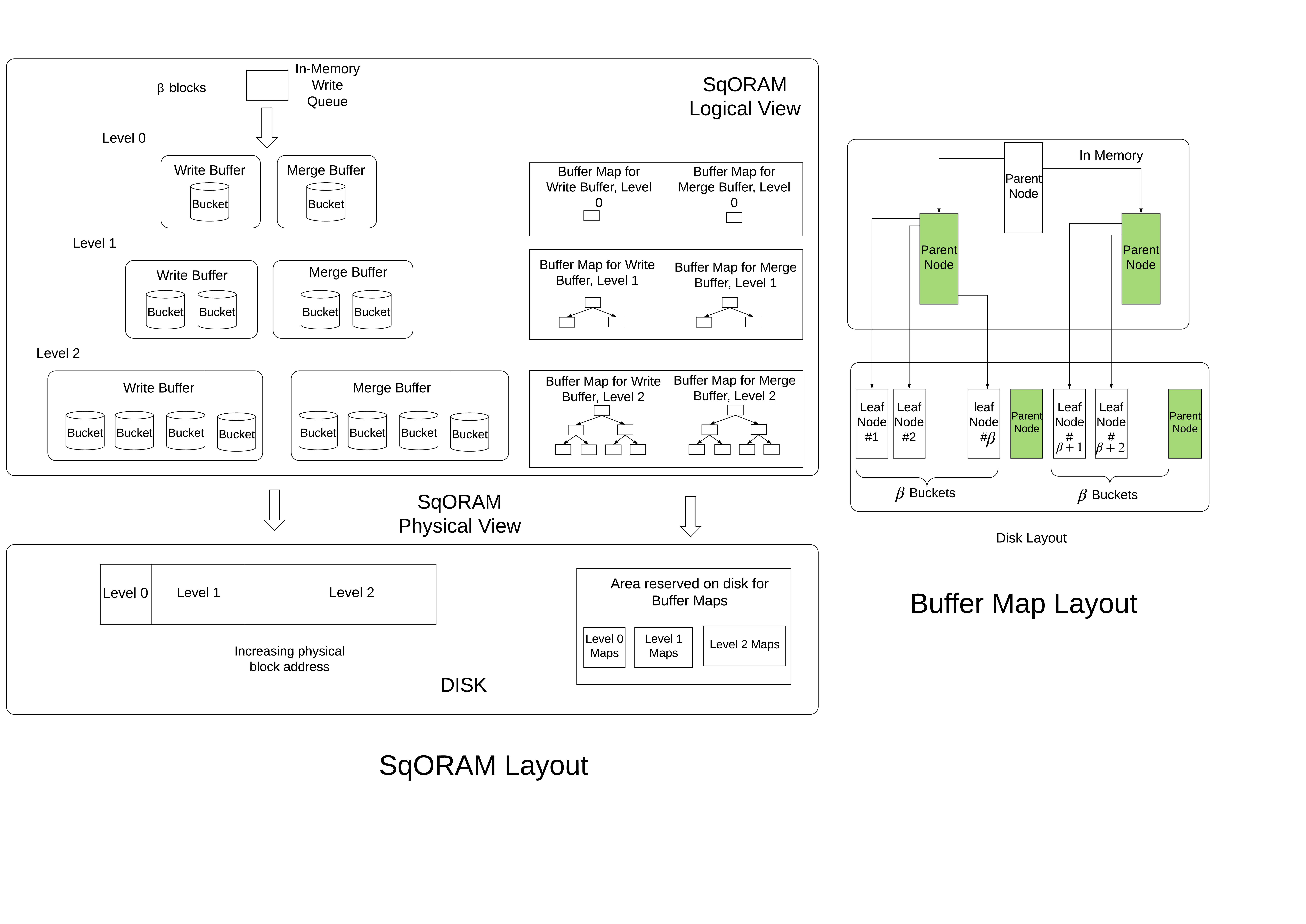}
	\vspace{-2cm}
	\caption{\small {\sys} Layout. The database is organized into  $\log_{k}(\frac{N}{B})$ levels. Each level contains two identical buffers. Each buffer in level $i$ contains $2^{i}$ buckets with $\beta$ data blocks. Levels are stored sequentially to disk. Each buffer has a {\em buffer map} (B-tree) to quickly
		map logical IDs to blocks within that buffer.\label{layout}}
	%\vspace{-1.0cm}
\end{figure*}

% {\color{blue}
{\sys} aims to perform locality-preserving reads and writes, while preserving write-access privacy. A good starting point for this is to place logically-related data close-together on disk initially when access are sequential in the logical domain e.g., by writing logically-related data blocks to adjacent physical blocks {\em sequentially} \cite{usna_stash_free_oram}. In an initial stage, the disk layout resembles an append-only log, where the {\em next} logical write is performed by writing data to the head of the log.  

The next critical task is to maintain the layout as data ends up being scattered across the disk over time \cite{betrfs}. One way to achieve this is by periodically reshuffling data to bring logically-related items in close proximity. Importantly, the frequency of reshuffles and the corresponding access patterns should not leak write-access privacy.

{\sys} adopts a physical layout similar to hierarchical ORAMs (Section \ref{sqoram:background:hierarchical_oram}) with several key differences. In this section we overview the {\sys} construction and present key insights. Further details are provided in later sections.
% }

\subsection{Locality-preserving disk layout}

\paragraph{Organization}
Similar to the case of hierarchical ORAMs, in {\sys} $N$ data blocks are organized in a ``pyramid'' with multiple {\em levels}, each level twice the size of the preceding one.  Levels are further subdivided into two identical buffers -- a {\em merge} buffer and a {\em write} buffer. Their function will be discussed shortly. The buffers comprise multiple logical ``buckets'', each bucket containing up to $\beta$ fixed-sized blocks. The level number determines the number of buckets in the buffers: a buffer in level $i$ contains $2^{i}$ buckets and overall, level $i$ contains $2^{i+1}\times \beta$ blocks in total. The last level buffers contain $N/\beta$ buckets and can hold all $N$ blocks. The total number of logical levels is $\O(\frac{\log(N/B)}{\log(k)})$.

\paragraph{Insight  1:  Locality-Preserving Storage Invariant}
To preserve logical domain access locality (e.g., reads by an overlying file system) it is desirable to physically store blocks sorted on their logical addresses.
%\todo{above you say that they are stored in increasing order of physical addresses -- now you are refering to logical addresses -- which is it? is it because FSes are submitting things with locality to us? or is this entire argument messed up competely?}. 

To achieve this, {\sys} replaces level-specific hash tables used in standard hierarchical ORAMs with a layout sorting blocks by logical addresses. This is allowable since write-only privacy does not require storing blocks at random locations as in standard hierarchical ORAMs -- since the adversary does not see reads and thus cannot link reads with writes. Before being written to disk, blocks are re-encrypted/randomized. 

{\sys} stores levels in their entirety on-disk under the following invariant: 

\begin{myquote}
	Blocks in level buffers are written to disk in ascending order of their logical addresses.
\end{myquote}	

As will be discussed shortly, this layout enables seek-efficient level reshuffles.

\paragraph{Level Index}
To efficiently track the precise location of blocks, {\sys} stores a search index for the merge and write buffer in each level. 
In a given level, each buffer has its own on-disk B-tree index, named \emph{buffer map}. Each buffer map node is stored in a physical block. Buffer map tree leaf node entries contain tuples of the form $\mathsf{\<laddr, paddr\>}$, where $\mathsf{laddr}$ denotes the logical address of a block and $\mathsf{paddr}$ is an offset within the corresponding buffer where the block currently resides. Entries are sorted on $\mathsf{laddr}$. 

%{\color{blue}
Each internal node entry corresponds to a child node and is a tuple of the form $\mathsf{\<child\_addr, child\_paddr\>}$. Specifically, for each child node, $\mathsf{child\_addr}$ is the value of the {\em lowest} 
logical address noted in the child node and $\mathsf{child\_paddr}$ is the physical address corresponding to the location of the child node on disk.

We remark that the buffer maps allow faster queries than a binary search over the sorted blocks in the buffers. This is because each B-tree node is stored in a disk block which must be read/written as a whole, and the tuples are small in size -- 16 bytes each assuming 8 byte logical and physical addresses. Thus, a large number of tuples can be packed into a single node (e.g., 256 entries for 4KB disk blocks), for  a B-tree with a large fan out and small depth. Queries can then be performed more efficiently and with less number of seeks than a binary search. As we show in Section \ref{seqoram:amortized}, the B-tree maps can also be constructed in a seek-efficient manner.
%}

%\todo{if they are sorted why do we need an index? can't we binary search easily?}. 
%identical\todo{do you mean identical?!?! are 
%these the search indexes? if they are identical why not just keep one only?} 
%Fake blocks are represented by sentinel values in the leaf node.\todo{completely unclear -- does a 10 year old understand this layout so far? Is this a simple search tree? what does ``similarly organized'' mean? what do the internal nodes contain? you suddenly introduced the fake blocks here!??!?! were they ever discussed above???!? what are they and how are they used? etc.}

\subsection{Asymptotically Efficient Level Reshuffles}
For hierarchical ORAM constructions, expensive level reshuffle mechanisms are primarily responsible for high performance overheads. 
In the {\sys} construction described thus far, level reshuffles would constitute obliviously sorting the combined contents in the merge buffer and the write buffer of a level based on random tags (e.g., logical address hashes), 
and writing the sorted contents to the next level. This is not only expensive in terms of access complexity (featuring an asymptotic complexity of $\O(N \log N)$) but also seek-intensive as blocks are read from and written to random locations. 

\paragraph{Insight 2: Oblivious Merge For Level Reshuffles}
Sorted layouts come with a significant additional benefit: the ability to obliviously and efficiently merge the write and the merge buffers in a level to create the next level during level reshuffles.
Obliviously merging is asymptotically faster (by a factor of $\O(\log N)$) than oblivious sorting. Importantly, such oblivious merges are also more seek-efficient. Obliviously sorting blocks on randomly-generated tags requires at least $\O(N \log N)$ seeks. In contrast, obliviously merging blocks from the sorted buffers requires only $\O(N)$ seeks, as the merge can be performed in a single pass. In fact, with a small constant amount of memory, it is possible to reduce seeks further (Section \ref{seqoram:amortized}).

\paragraph{Worst-Case Construction}
Typically, hierarchical ORAMs \cite{goldreich}, amortize the cost of the expensive reshuffles over multiple queries. Unfortunately this comes with often prohibitive worst-case delays when clients need to wait (up to hours or days) for reshuffles to complete. This is often impractical for existing system stacks with pre-defined timeouts, such as file systems.

Several solutions have suggested de-amortizing the construction \cite{goodrich_deamortized,kushilevitzoram,privatefs}. However, as noted in \cite{privatefs}, these solutions do not {\em strictly} de-amortize the level reshuffle, since the subtasks involved in oblivious sorting have widely different completion times. Proper monitoring and {\em strict} de-amortization of hierarchical ORAMs is a non-trivial task.

Benefiting from the oblivious merges, {\sys} presents a naturally un-amortized construction where exactly the same amount of work is done per query, not unlike efficient tree-based ORAM designs \cite{pathoram}.

\subsection{Efficiently Tracking Blocks for Queries}
In hierarchical ORAMs, the exact location of a block is precisely determined by its last access time. Once a block is written to the top level, it moves down the levels according to a precise periodic level reshuffle schedule. Typically, during a query, each of the $\log N$ levels is searched for the matching block. Once found at a particular level, the search continues in the next levels by reading dummy blocks, to hide the location where the block has been found. However, when reads cannot be observed by the adversary, the search can stop as soon as the block is found at a particular level.

\paragraph{Insight 3: Tracking Block Locations Using Last Access Time Based Position Map}
Moreover, using the \emph{last access time} of a block, {\sys} can precisely track its location and perform queries efficiently by directly reading the block from the level where it currently resides. Time is measured by a write counter tracking the number of writes since initialization. Last access time information, in conjunction with the current time (value of write counter = total number of writes since initialization), allows a precise determination of the level and buffer in which a particular block resides. 

The critical challenge is to privately and efficiently store this information. With enough in-memory storage, the last access times can be stored in memory, and synced to the disk on clean power-downs. On power failure or dirty power-downs, the information can be reconstructed in a linear pass over the level indexes only.
%\todo{is this correct?} 

However, with limited memory, this information needs to be stored and obliviously queried from disk. To this end, {\sys} maintains an oblivious \emph{access time map} (ATM) structure. The ATM is similar to a B+ tree with one key difference -- {\em instead of each B+ tree node storing physical addresses as pointers to its children, the last access times of the children nodes act as pointers and are stored in the nodes. }
 ATM nodes are stored in the same ORAM along with the data. 
The ATM can be traversed from the root to the leaf for determining the location of each child node on the path in the ORAM, based on its last access time value. This is detailed in Section \ref{deamortized:atm}. 
%\todo{it is traversed by determining? or for determining? or what? confusing sentence}

Using the ATM, {\sys} can reduce the number of index lookups during queries by a factor of $\O(\log N)$.
% (since all levels do not need to be searched).

\section{Amortized Construction}
\label{seqoram:amortized}

We first introduce an amortized construction to demonstrate our key idea. Later Section \ref{seqoram:deamortized} shows how to de-amortize efficiently.

\paragraph{In-memory storage}
In addition to disk storage, {\sys} has memory sufficient to hold a constant ($c$) number of buckets.  This is used for performing in-memory bucket merges during level reconstructions.

\paragraph{Search Invariant}
As with most hierarchical ORAM constructions, {\sys} ensures the following invariant:

\begin{myquote}
The most recent version of a block is the first one found when ORAM 
levels are searched sequentially in increasing order of their level number. 
\end{myquote}

\subsection{{\sys} Operations}
{\sys} supports three operations:

\begin{itemize}
	\item {\em $\mathsf{write(b,d)}$:} Writes block with address $b$ and data $d$.
	\item {\em $\mathsf{merge(i)}$:} Merge contents of the buffers in level $i$ and write to level $i+1$ on disk.
	\item {\em $\mathsf{read(b)}$:} Read block with address $b$ from the ORAM.
\end{itemize}

\paragraph{Writes}
{\sys} performs data block writes to an in-memory {\em write queue}.  The
queue is of the size of a bucket. When the
write queue is full (after $\beta$ writes), its blocks are sorted on their logical block addresses and flushed to the {\em write} buffer of the ORAM
top level.

%%\smallskip

\paragraph{Merging Levels: Intuition}
Once the contents of the write queue has been written to the write buffer of the top level, 
{\sys} checks the state of the merge buffer of the top level. Specifically, at this stage the following two cases are possible: 
\begin{itemize}
	\item {\em Merge buffer is empty:} In this case, the buffers are logically switched -- the write buffer becomes the merge buffer and the previously empty merge buffer becomes the write buffer for future accesses. 
	
	\item{\em Merge buffer is full:} In this case, the contents of the write buffer and the merge buffer of the top level are merged together to create the write buffer of the second level. To this end, the (sorted) write buffer and merge buffer buckets are read into memory. The two buckets are merged, their blocks re-encrypted and written sequentially to the write buffer buckets in the second level.  
	
\end{itemize}

%Figure \ref{merge_example} presents a demonstrative example of the level merge procedure for the first two levels. 

\paragraph{Merging Levels: Protocol}
Formally, $\mathsf{merge(i)}$ (Algorithm \ref{merge_level} in Appendix) includes the following steps:

\begin{itemize}
	\item {\em Setup:} Initialize two {\em bucket-sized} queues in memory corresponding to the write buffer ($q_w$) and the merge buffer ($q_m$) of level $i$.
	
	\item {\em Fill up queues:} Read sequentially from the corresponding buffers to the respective queues until full (Steps 2 - 13). In case all $2^{i}\cdot \beta$ blocks have already been read from the buffers, {\em fake} blocks (blocks assigned a logical block address of $N+1$, containing random data) are added to the queues instead.
	
	\item {\em Write blocks from queues to next level until empty:} Retrieve a block {\em each} from both the queues, compares 
	their logical addresses and writes back the 
	block with the lower logical address {\em sequentially} to the write buffer in level $i+1$ (Steps 14 - 25).

\end{itemize}

%In general (Algorithm\ref{merge_level}), once all the write buffer and merge buffer buckets in level $i-1$ have been filled up, they are merged and written to the write buffer of level $i$. Note that at this stage, level $i-1$ contains $2^{i}\beta$ blocks in total -- equal to the size of the write buffer of level $i$. 
%
%{\sys} initializes two {\em bucket-sized} queues in memory corresponding to the write buffer ($q_w$) and the merge buffer ($q_m$) of level $i-1$. Then, blocks are read sequentially from the corresponding buffers to the respective queues (Steps 2 - 13). {\em In case all $2^{i-1}\beta$ have already been read from either of the buffers, {\em fake} blocks (blocks assigned a logical block address of $N+1$, containing random data) are added to the queues instead.}
%
%Next, {\sys} retrieves a block {\em each} from both the queues, compares 
%their logical addresses and writes back the 
%block with the lower logical address {\em sequentially} to the write buffer in level $i$ (Steps 14-25). This is continued until the write buffer of level $i$ is full ($2^{i}$ writes). 

\paragraph{Handling Duplicates}
If the queues contain duplicates then 
the block in the write buffer queue($q_w$) will be written (as it is more recent) and the block in the merge buffer queue ($q_m$) will be discarded (Step 21-22). Also, since {\em fake} blocks have a logical address of $N+1$, 
real blocks will be written to level $i$ before {\em fake} blocks. This however is not a security leak since 
the location of fake and real blocks are important only while performing reads -- which is 
not protected by {\sys}. Semantic security ensures content-indistinguishability.

%While creating the last level (Steps 26 -37), the writes are performed in a slightly different fashion. Specifically, if the next write to the buffer in the last level is to offset $ctr_{next}$, {\sys} checks whether either of the blocks retrieved from the queues have logical address equal to $ctr_{next}$. If not, {\sys} reencrypts the block already at that offset (Step 27 - 28). Otherwise, a block is written from the queues as described before (Step 30 - 37). This ensures that the offset at which a given block is written to the last level is equal to its logical address.

%Note that merging and creating levels as described above does not require reading and storing the merge buffer and the write buffer of level $i$ in its entirety in the memory at a given time. Instead, memory is used only for storing the two bucket-sized queues ($q_w$ and $q_m$). 

\paragraph{Bottom-up Buffer Map Construction}
{\sys} employs a novel mechanism to optimize the number of disk seeks that need to be performed to construct a particular B-tree buffer map. Specifically, the B-tree buffer map for the write buffer of a level is constructed in a {\em bottom up} fashion when blocks are written 
as a result of a merge (Algorithm \ref{merge_level}, Step 24). In particular, after a new bucket is written to the write buffer of a level (Step 14 - 25, Algorithm \ref{merge_level}), a new leaf node is added to the corresponding B-tree buffer map, 
containing logical and physical addresses of blocks in that bucket. 

After $\beta$ leaf nodes have been added to the B-tree as a result of subsequent accesses, a parent node of the  $\beta$ leaf nodes is created 
in memory with an entry for the {\em minimum} of the logical block addresses in the corresponding 
leaf nodes (Figure~\ref{layout}). Once the parent node has $\beta$ entries, the parent node is written to the 
disk as well. In this way, parent nodes are created in memory before being written to the disk next to the children nodes. {\em Writing parent nodes next to the children nodes sequentially on disk optimizes the number of seeks that need to be performed while constructing the tree.} 
Note that this requires $\O(\frac{\log N}{\log \beta})$ blocks 
of memory in order to store the parent nodes up to the root.

\BT
	\label{thm:amort_correctness}
	The amortized merge procedure (Algorithm~\ref{merge_level}) ensures that all data blocks from the merge and write buffers of level $i-1$ (for any $i \leq logN$) are merged in ascending order of their logical addresses and written to level $i$, within $2^{i}\cdot \beta$ accesses.
\ET

\begin{proofsketch}
The result is based on the following:
\begin{itemize}
	\item There are at most $2^{i}\cdot \beta$ real data blocks in level $2^{i}$ spread across the merge and write buffers. 
	\item Fake blocks (with address $N+1$) are added to the in-memory queues only after all data blocks in the merge/write buffer have been written to level $i+1$.
	\item Data blocks in either of queues are written to level $i+1$ before fake blocks.
\end{itemize}	
	
The full proof is detailed in the Appendix. 

\end{proofsketch}

\begin{theorem}
	\label{thm:amort_security}
The amortized merge protocol (Algorithm~\ref{merge_level}) ensures that during the merge all writes to level $i$ are uncorrelated and indistinguishable, independent of the logical block addresses..
\end{theorem}

\begin{proofsketch} 
It is straightforward to see that while merging the buffers of levels $i-1$ and writing to level $i$ (for any $i \leq logN $), the only step observable to the adversary is Step 24 (Algorithm~\ref{merge_level}). The rest of the steps involve reads or in-memory operations.

Step 24 performs a write to the {\em next} block in sequence to the write buffer in level $i$, irrespective of the logical block address and content. Also,  invariably the the process is continued until $2^{i}\cdot \beta$ blocks have been written (Step 1). In case, the number 
of real data blocks is less than $2^{i}\cdot \beta$, {\em fake} blocks are written instead. Semantic security ensures that fake blocks are indistinguishable from real data blocks.

\end{proofsketch}

\begin{figure}[t!]
	\centering
 \includegraphics[scale=0.1]{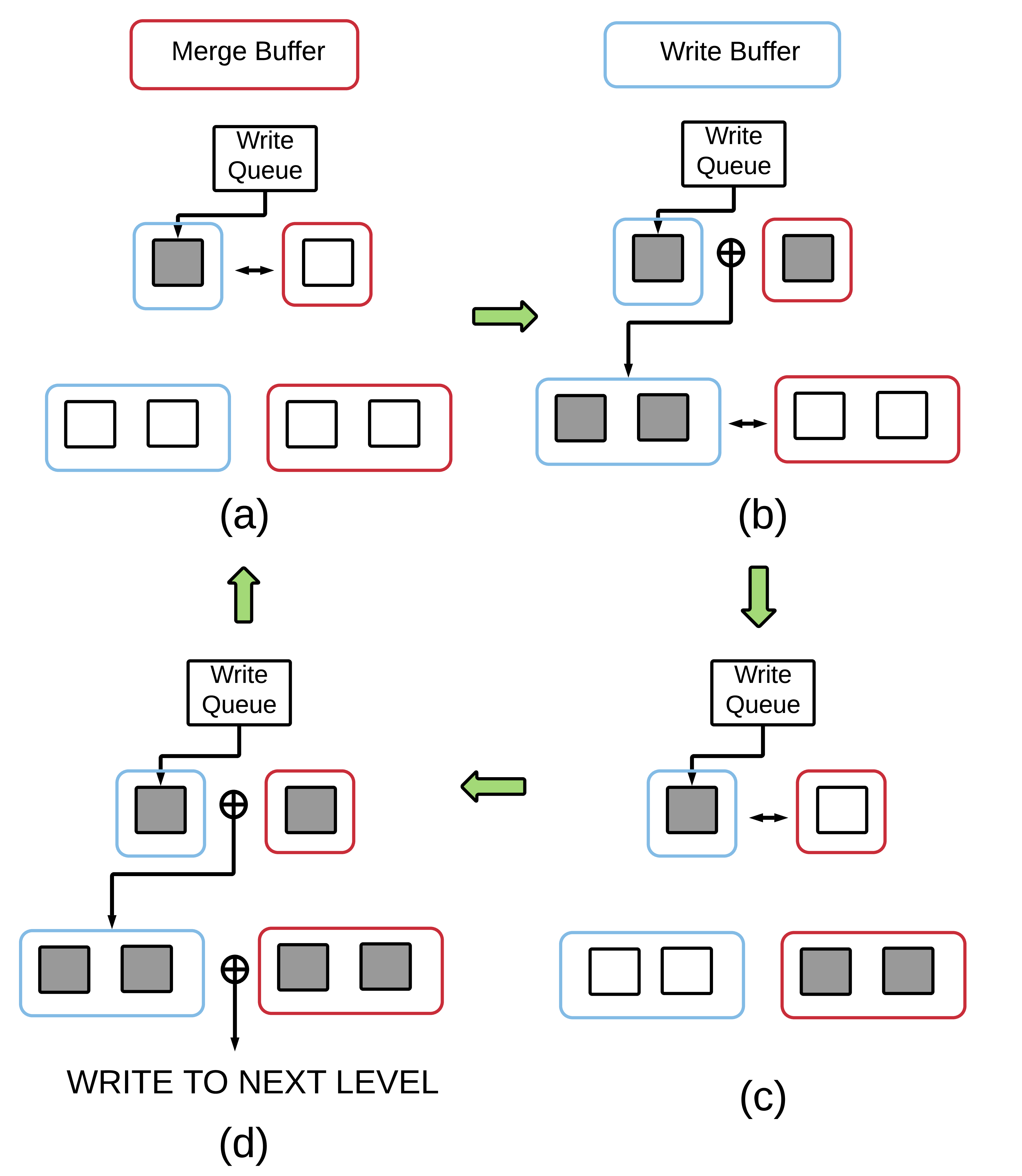}
 \vspace{-0.3cm}
 \begin{footnotesize}
 \caption{{\small Illustrative example of the merge protocol. (a) blocks in the write queue are written to the write buffer in level 0. (b) the write queue is flushed to the empty write buffer in level 0. After this, the blocks in the write buffer and merge buffer of level 0 are merged and written to level 1. Once the write buffer of level 1 is full, the buffers in level 1 are switched. Figure (c) and (d) show how the two buffers in level 1 are similarly filled and then merged and written to the next level(s).  }\label{merge_example}}
 \end{footnotesize}
\end{figure}

%\begin{figure}
%\centering
% \includegraphics[scale=0.16]{figures/map_tree_layout.eps}
% \vspace{-1.4cm}
% \begin{footnotesize}
% \caption{
%Layout of the map B+ tree on the disk. While $\beta$ leaf nodes are written sequentially 
% to the disk, a parent node is created in memory and updated with the value of the minimum logical block address 
% per leaf node. Once the parent node has $\beta$ entries, it 
% is written to the disk sequentially after the leaf nodes. \label{map_tree_layout}}\end{footnotesize}
%\end{figure}

\paragraph{Reads}
{\sys} reads are similar to queries in hierarchical ORAM 
constructions. Specifically, reads are performed by searching each level in the ORAM {\em sequentially} for the required block. A block may reside either 
in the write buffer or the merge buffer of a particular level at any given time. Therefore, both the buffers in a level must be checked for a block -- \emph{{\sys} checks the write buffer first since it contains blocks that have been more recently updated than the blocks in the merge buffer. As a result, the most up-to-date version of a block is found before other (if any) invalid versions.}

Retrieving a particular block requires querying the maps for the write buffer and the merge buffer (in order) 
at each level, starting from the top level and sequentially querying the levels in increasing order of the level numbers, until an entry for the block is found. 
Then, the block is read from the corresponding buffer of the corresponding level. Observe that only the required block is read and not the entire bucket since the location of a block in a bucket can be precisely determined based on the sorted order of blocks.

\paragraph{Write Access Complexity}
Note that during construction of level $i$, $2^{i-1}$ buckets each in the write buffer and the merge buffer in level $i-1$ are merged and written to level $i$. For the merge $2^{i}$ buckets in total have to be read from level 
$i-1$ {\em exactly once} while $2^{i}$ buckets are 
written {\em exactly once} to level $i$. Further, 
constructing the B-tree map for the write buffer in level $i$  requires writing $2 \times 2^{i}$ blocks. Level construction for level $i$ can thus be performed with $\O{2^{i}}$ 
accesses. 

Since each level is exponentially larger than the previous level, level $i$ is constructed only after $2^{i} \times \beta$ writes. The amortized write access complexity is:

\begin{center}
	$\sum\limits_{i=0}^{log_{2}{N/B}} \frac{\O{2^{i}}}{2^{i} \times \beta} =\O(\log{N})$
\end{center}

\paragraph{Read Access Complexity}
The read access complexity of the amortized construction is $\O({\log{N} \times \log_{\beta}{N}})$ 
since to read a block, a path in the B-tree buffer maps at each level must be traversed to locate the level at which the block exists. 
%
%
%Since, each buffer map leaf node is written to an individual physical block, and 
%each entry in the leaf node is of a fixed size (since logical and physical block address 
%sizes are fixed), the number of entries that can fit in a leaf node can be precisely determined. 
%Let $B$ be the physical block size and $|addr|$ the size of an address (tuple in the leaf 
%node). Then, sizing a bucket to contain $\beta = B/|addr|$ blocks ensures 
%that all these blocks have entries within the same leaf node (of size $B$). 
%
Each buffer in level $i$ has $2^{i} \times \beta$ blocks. To determine the height of the 
buffer map B-trees for a level, note that each leaf of the tree contains $\beta$ tuples. With each 
tuple then corresponding to a block, the number of leaves in the B-tree for that level is $2^{i}$.
Consequently, the height of the B-trees (with fanout $\beta$) for level $i$ is $\log_{\beta}{2^{i}} = \O(\log_{\beta}{N})$. 
For a 1TB disk with 4KB blocks and 64-bit addresses, $\beta = 256$ and $\log_{\beta}{N} = 4$. 
%In later sections we reduce the read complexity.

\BT[Seek analysis]
The number of disk seeks performed by {\sys} for level reshuffles across $\log{N}$ levels, where $N$ is the number of blocks in the ORAM, amortized over the number of writes is $\frac{4\log{N}}{ \beta}$.
\ET

\begin{proofsketch}
	The result is based on the following: 
	
	\begin{itemize}
		\item Reading $\beta$ blocks to each of the in-memory queues requires two disk seeks -- to place the read head at the starting location of the write buffer and then the merge buffer.
		\item After the queues are full, $\beta$ blocks are written to the write buffer of level $i$ {\em sequentially}. This requires {\em one} disk seek after which the queues are again filled up.
		\item A leaf node is added to a buffer B-tree map which requires only one disk seek. 
	\end{itemize}	
	
The full proof is detailed in the Appendix.
\end{proofsketch}

 {\em This is an important result -- if the bucket size is $\O({\log{N}})$, the amortized number of seeks required for the level reshuffles in {\sys} is a constant.} A bucket size of $\O({\log{N}})$ entails allocating 
 $\O({\log{N}})$ blocks of memory for storing the queues. This is not impractical -- e.g., the actual memory required to be allocated for the queues in order to achieve an amortized number of seeks equal to 1,
  with 4KB blocks and 1TB disk is $M = 2 \times 4 \times (30) = 240$ blocks, or 960KB. 
 
% In fact, to reduce seeks even further, more memory can be allocated in memory-rich systems. Increasing the memory allocated reduces the number of seeks required for the merge since with larger queues, a large part of the buffers can be read {\em sequentially} without incurring a seek in between.

% Observe here that the seek analysis implicitly assumes that all level reshuffles are performed {\em atomically} -- no queries are performed in between. Indeed, if block queries (reads) are interleaved with the level reshuffles, the number of seeks will also include the cost of the read in between, as disk I/O cannot be performed in parallel. This will invariably increase the number of disk seeks.
% 
% {\sys} can support reads in between level reshuffles because even when the write buffer of a level is partially constructed, the buffer map required to retrieve blocks from the (partially created) buffer is constructed simultaneously. This idea is used to support interleaving reads even in the de-amortized construction (Section \ref{seqoram:deamortized}).
% 

 \begin{theorem}
  The amortized {\sys} construction provides write-access privacy (Definition \ref{def:privacy}).
 \end{theorem}
 
\begin{proofsketch}
The proof follows straighforwardly by construction.  Consider two equal length
write access patterns, $\vec{A} = {w_1, w_2, \ldots w_i}$ and $\vec{B} = {x_1, x_2, \ldots
x_i}$. When either of the access patters are executed, $i$ encrypted blocks are first added to the in-memory write queue irrespective the logical addresses.  Once the write queue is full, its contents are written the top level encrypted with semantic security. The top level contents do not leak any information about whether $\vec{A}$ or $\vec{B}$ was executed.

Flushing the write queue will trigger level reshuffles for $k < \log N$ levels. Theorem \ref{thm:amort_security} shows that the writes to the disk while reshuffling any level are uncorrelated to each other and independent of the block addresses and content.  Therefore, 
the writes performed for reshuffling level $j \leq k$ when $\vec{A}$ is indistinguishable from the writes performed when $\vec{B}$ is executed. 

Further, level reshuffles are independent of each other and are triggered at periodic intervals based on only the number of writes performed (public information). Therefore, by observing the writes to the top level and the writes due to the level reshuffles an  adversary can only do negligibly better than purely guessing whether $\vec{A}$ or $\vec{B}$ was executed.
%
%
%
%level
%reconstructions are triggered independent of the contents of the write queue.
% 
% Theorem \ref{thm:amort_security} shows that the level reshuffles for any level $i \leq \log N$ are performed while ensuring that the access transcripts do not leak any information about the blocks that are being reshuffled. .  Thus, when either of $A$ and $B$ are executed, the access transcripts for the $i$ writes and the corresponding level reshuffles that this triggers will be indistinguishable. Observing these access transcripts, 
%%  
%During the level reconstructions, blocks from two buckets are merged
%together on the basis of their logical addresses.  Since reads are not
%observable, the buckets that are being merged are not known to the
%adversary.  The order in which the merged buckets are written is independent
%of the result of the merge.  More specifically, during a merge for level
%$i$, an adversary observes buckets always written in the order of 1 to
%$2^{i}$ irrespective of the contents of the buckets.  Since, blocks are
%written reencrypted during a merge, the semantic security of the encryption
%scheme provides indistinguishability between the blocks written to a bucket
%after the merge.  
% 
\end{proofsketch}

\section{Deamortized Construction}
\label{seqoram:deamortized}

\begin{figure}
\centering
 \includegraphics[scale=0.20]{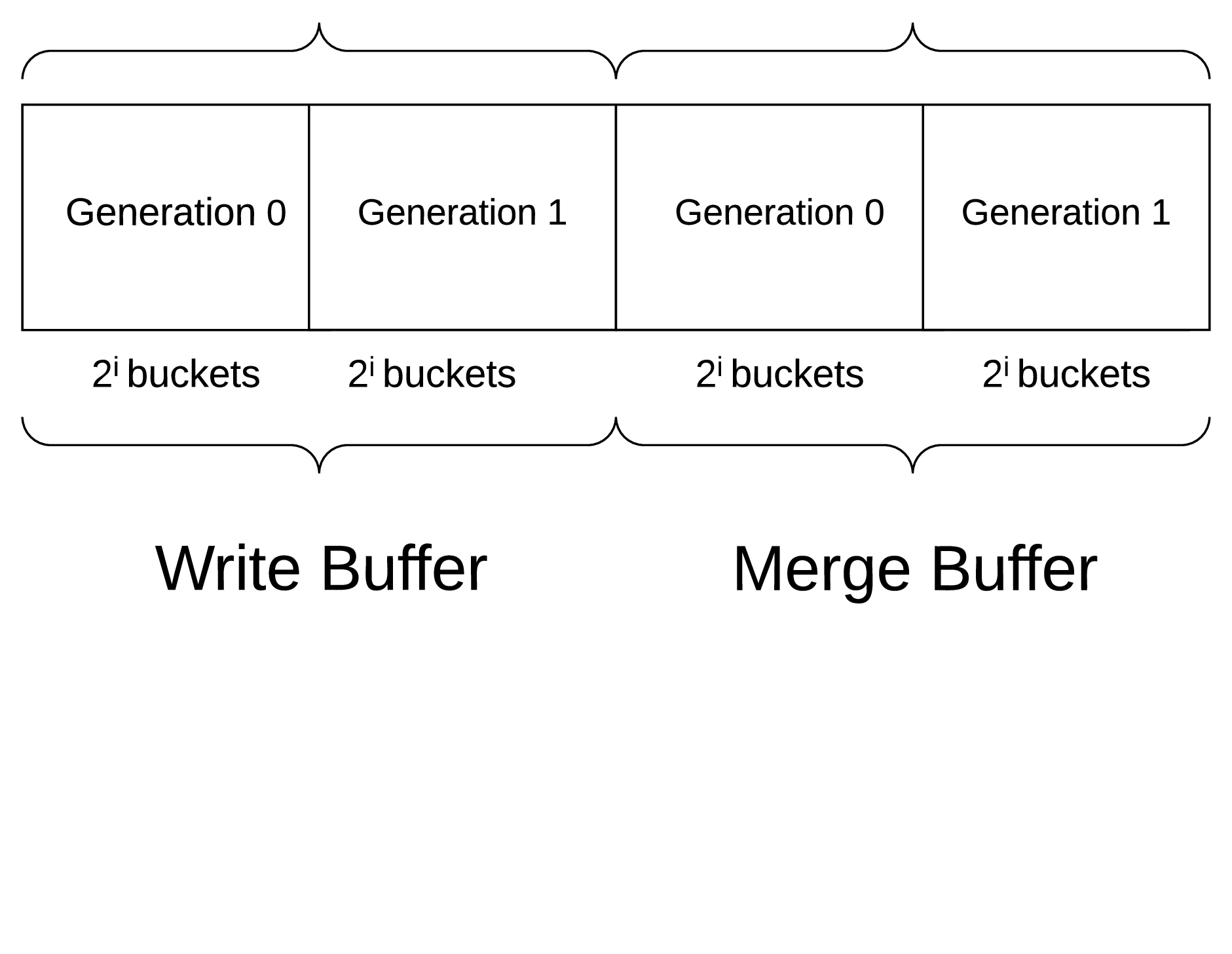}
 \vspace{-1.8cm}
\caption{\small Level design for the de-amortized construction. Level $i$ has two 
sets containing $2^{i}$ buckets for each buffer. The buckets in the two sets are denoted as {\em generation 0} buckets and {\em generation 1} buckets respectively \label{level_design}}
 \end{figure}
 
The amortized construction achieves 
appreciable performance incentives over \cite{hive} by reducing the amortized write access complexity and 
number of seeks per write. However it also suffers from two major drawbacks: i) the read access 
complexity is higher and ii) the worst case write access complexity is $\O(N)$ (for merging and reconstructing the last level).  
%In general, due to the high worst case costs, amortized constructions perform poorly on existing systems. For instance, in order to achieve constant seek time guarantees, level reconstruction, especially for larger levels, 
%will block upper level I/O requests for significant periods of time. This will violate standard
%I/O timeout parameters  and make the system impractical
%
Therefore, to make {\sys} usable in practice, we first present a practical de-amortized version in the following and then describe how to reduce the read access complexity.

\subsection{Deamortized Writes}
\label{deamortized:writes}
De-amortization for hierarchical ORAMs is achieved by leveraging extra space~\cite{goodrich_deamortized,kushilevitzoram,privatefs}.
These techniques effectively ensure that each query takes roughly the same amount of time by monitoring progress over subtasks and forcing query progress to be proportional to level construction~\cite{privatefs}.
However, as noted in \cite{privatefs}, this does not {\em strictly} de-amortize the level reshuffle, since subtasks have widely different completion times -- correct monitoring and {\em strict} de-amortization of hierarchical ORAMs is a non-trivial task.

{\em In contrast, our key idea is to leverage the fact that {\sys} does not protect reads and achieve {\em strict} de-amortization.} This ensures that each write performs exactly the same amount of work and has identical completion time, eliminating the need for additional synchronization between queries and reshuffles.

%\begin{algorithm}
%	\caption{Initialization}\label{merge_init}
%			\begin{footnotesize}
%		\begin{algorithmic}[1]
%	\State $q_{x0} := \phi$ // Persistent in-memory queue of size $\beta$ assigned for level $x < \log{N}$
%	\State $q_{x1} := \phi$ // Persistent in-memory queue of size $\beta$ assigned for level $x < \log{N}$
%	\State $ctr_{x0} := 0$ // Persistent in-memory counter assigned for level $x < \log{N}$  
%	\State $ctr_{x1} := 0$ // Persistent in-memory counter assigned for level $x < \log{N}$ 
%	\State $g := 0 $ // Global Access Counter (number of times write queue has been flushed to the disk)
%\end{algorithmic}
%\end{footnotesize}
%\end{algorithm}

\paragraph{Key Differences from Section \ref{seqoram:amortized}}
In order to de-amortize the construction we make several changes: 

\begin{itemize}
	\item {\em Leveraging extra space to continue reshuffles in background with queries:} 
	As in ~\cite{goodrich_deamortized,kushilevitzoram}, {\sys} uses extra space per level 
	to continue writes while reshuffling. In particular, each bucket in a level is duplicated -- 
	the two set of buckets are termed {\em generation 0} and {\em generation 1} buckets respectively (Figure \ref{level_design}). Each generation is augmented with a B-tree search index (similar to the buffer maps in the amortized construction)
	
	\item {\em Merging contents in generations:} 
	Instead of merging blocks in the merge buffer and the write buffer of a level, the blocks in the generation 0 and generation 1 buckets of the merge buffer are merged together and written to the write buffer of the next level. The results of the merge are written to  a particular generation of the  write buffer in the next level:
	\begin{itemize}
		\item If generation 0 buckets are empty, then the blocks after the merge are written to the  generation 0 buckets.
		\item If generation 0 buckets are already full, the blocks after the merge are written to the  generation 1 buckets. 
	\end{itemize}
	Once the write buffer of a level is full, it is switched with the merge buffer. At this stage, the merge buffer is  invariably empty. Buffers are used alternatively for merging levels (merge buffer) or for 
	writes from the previous level (write buffer). 
\end{itemize}

%%\smallskip
\paragraph{Last Level Organization}
In addition, the last level is organized differently from the other levels -- the last level contains only one buffer with $N/\beta$ buckets, or $N$ blocks in total. Blocks in the last level are also placed in a different manner --  the offset at which a block is placed in the last level buffer is determined by its logical block address -- e.g., if the logical block address of a given block is $l$ and the last level buffer starts from physical address $x$, the block will be placed at physical address $x+l$. Contrast this with other levels where blocks within a buffer are sorted according to the logical address but the physical location has no correspondence with the logical address. 
As we will show later, this does not leak security and is crucial for the correctness of our de-amortized merge protocol.

\begin{figure*}[th!]
\centering
\includegraphics[scale=0.15]{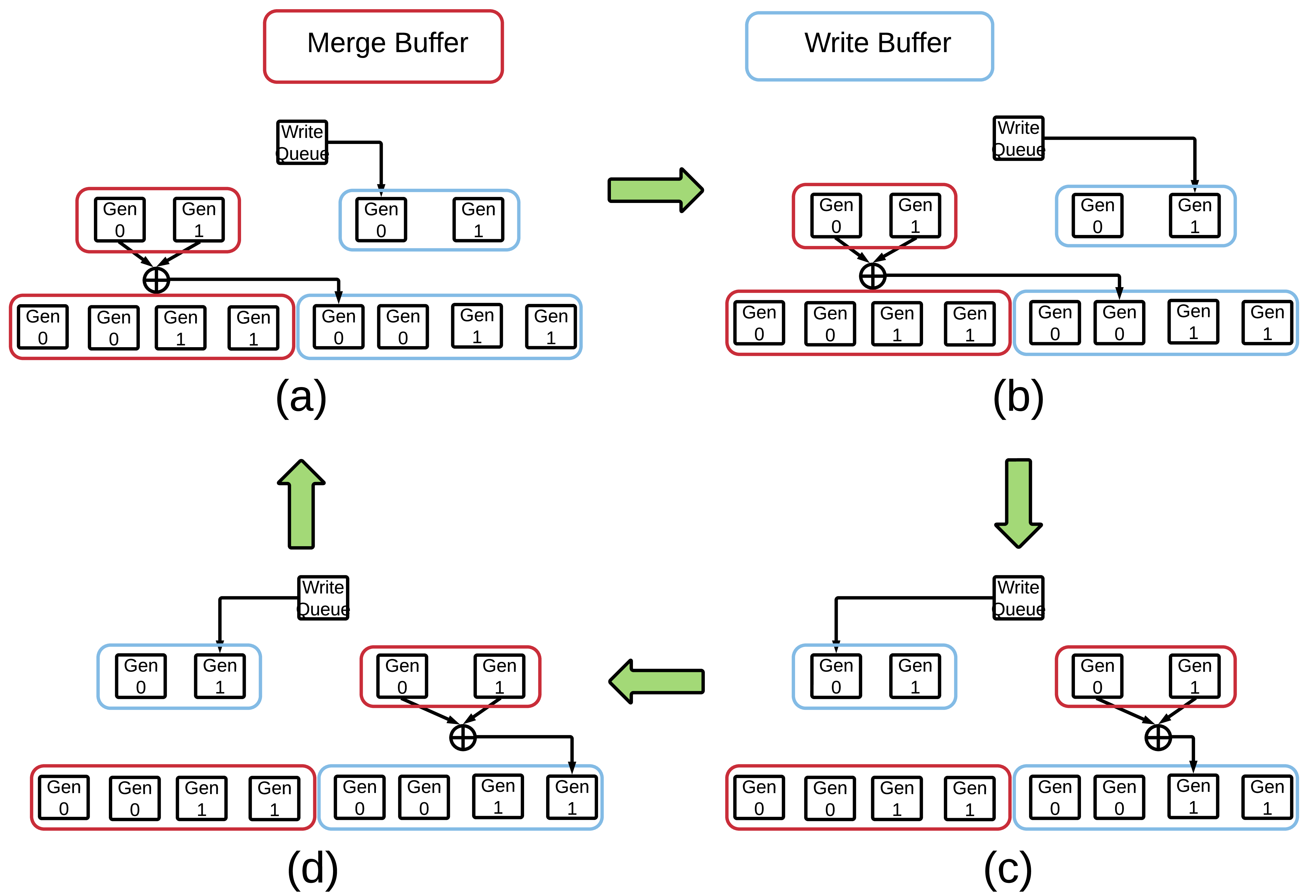}
%\vspace{-0.4cm}
\caption{\small De-amortization example for 3 levels. In (a) and (b), the generation 0 and generation 1 buckets in the merge buffer of 
level 1 are merged to form generation 0 of level 2 while writes from the write queue are 1 to the 
write buffer in level 1. Once generation 0 in write buffer in level 2 has been written (and the merge in 
the merge buffer of level 1 has been completed), the buffers are switched. In (c) and (d), the merge in 
level 0 creates generation 1 of level 2 while writes are performed to the write buffer.
\label{deamor_ex}} 
\end{figure*}

%Figure \ref{deamor_ex} illustrates the de-amortized merge protocol.

%%\smallskip
\paragraph{De-Amortized Merge: Intuition}
%
%The de-amortized merge protocol writes $\beta$ blocks {\em sequentially} to a bucket in each of the $\log{N}$ levels. The specific bucket that the blocks are written to in a particular level is determined based on the current value of the global access counter. 
%
%In particular, since the write buffer in level $i$ contains $2^{i+1}$ buckets in total (two generations with $2^{i}$ buckets each) and one new bucket is written each time a merge is executed after a write queue flush, all $2^{i+1}$ buckets in the write buffer will be written once the write queue has been flushed $2^{i+1}$. After that, the write buffer is switched with the empty merge buffer and the next bucket to be written to the new write buffer in level $i$ will be the {\em first} bucket. 
%
%
%
%
%
%
%
%
%After every write queue flush to the top level write buffer, the merge protocol is executed. {\sys} tracks the number of times the write queue has been flushed to the disk during initialization using the {\em global access counter} $g$. 
%
{\em Essentially, the de-amortized merge protocol writes $\beta$ blocks {\em sequentially} to a bucket in each of the $\log{N}$ levels. }
The specific bucket that is written at a particular level is determined based on the current value of the global access counter. 
%-- $ctr_{next}$ simply determines the starting location of the bucket as on %offset in the buffer. 
In particular, since the write buffer in level $i$ contains $2^{i+1}$ buckets in total (two generations with $2^{i}$ buckets each) and one new bucket is written each time a merge is executed after a write queue flush, all $2^{i+1}$ buckets in the write buffer will be written once the write queue has been flushed $2^{i+1}$. After that, the write buffer is switched with the empty merge buffer and the next bucket to be written to the new write buffer in level $i$ will be the {\em first} bucket. {\em The de-amortized merge protocol essentially reconstructs each level in tandem, one bucket at a time.}

%Figure~\ref{deamor_ex} shows a demonstrative example of the de-amortization technique for an ORAM with 3 levels. 

\paragraph{De-Amortized Merge: Protocol}
Formally, the de-amortized merge protocol, $\mathsf{merge\_deamortized}$ (Algorithm \ref{merge_level_deamortized} in Appendix) includes the following steps:

\begin{enumerate}
	\item {\em Setup:} The de-amortized merge protocol requires few supporting counters that are initialized during setup and two queues for each level:  
	\begin{itemize}
		\item $q_{x0}$ -- In-memory queue of size $\beta$ used to store blocks from \emph{generation 0} of the merge buffer of level $x < \log{N}$.
		\item $q_{x1}$ -- In-memory queue of size $\beta$ used to store blocks from \emph{generation 1} of the merge buffer of level $x < \log{N}$.
		\item $ctr_{x0}$ -- Tracks the number of blocks that have already been read to $q_{x0}$ from \emph{generation 0} of the merge buffer 
		of level $x < \log{N}$.
		\item $ctr_{x1}$ -- Tracks the number of blocks that have already been read to $q_{x1}$ from \emph{generation 0} of the merge buffer 
		of level $x < \log{N}$.
	\end{itemize}
	
	\item {\em Merge buffers and write to next level:} For each level $x < \log N$, perform the following sub-steps -- 
	
	\begin{enumerate}
	\item {\em Fill Up Queues:} Fill up the two queues $q_{x0}$ and $q_{x1}$ with blocks read {\em sequentially} from the generation 0 buckets and generation 1 buckets of the merge buffer in level $x$ (Lines 6 - 17). Similar to Algorithm \ref{merge_level}, 
	if all real blocks have been read from the two generations, {\em fake} blocks are added instead (lines 8, 14). The values of $ctr_{x0}$ and $ctr_{x1}$ indicate the next blocks to be read from the respective generations. 
	
	\item {\em Write to next level:} Write $\beta$ blocks {\em sequentially} to the write buffer in level $x+1$ (Lines 18 - 31).
	\end{enumerate}
	 
\end{enumerate}

\paragraph{Writing to the Last Level}
The merge to the last level is handled slightly differently. Recall that due to the special organization of the last level, a block with logical address $j$ must be invariably written to offset $j$ within the last level buffer. 
Thus, the value of $ctr_{next}$ for the last level is determined keeping in mind that there is only a single buffer with $N/\beta$ buckets in the last level (Line 5). $ctr_{next}$ points to the next offset in the last level where the next block will be written after the merge. In case this does not match with the logical address of either of the blocks from the two previous level queues, the block at that offset is re-encrypted for indistinguishability (Lines 31 - 32). Otherwise, the required block is written from one of the queues assigned for the second to last level. (Line 34 - 41). 
%In case of the last level, the write is to the {\em write buffer} (Line 28).

\begin{theorem}
	\label{thm:deamort_correctness}
	The de-amortized merge protocol (Algorithm~\ref{merge_level_deamortized}) ensures that by the time the write buffer of level $i < \log{N}$ is full, all real data blocks from the merge buffer of level $i$ have been written to level $i+1$.
\end{theorem}

%\begin{proofsketch}
%Every successive execution of Algorithm \ref{merge_level_deamortized} writes $\beta$ blocks to the write buffer of level $i$ {\em sequentially}. At the same time, $\beta$ blocks are written from the merge buffer of level $i$ to the write buffer in level $i+1$. Note that within exactly $2^{i+1}$ successive executions of Algorithm \ref{merge_level_deamortized}, the write buffer of level $i$ will be full. But within the same time, $2^{i+1}\cdot \beta$ blocks will have been written from the merge buffer of level to level $i+1$ (Line 18 - 29).
%From Theorem \ref{thm:amort_correctness}, all real blocks from the merge buffer of level $i$ will necessarily be written to level $i+1$ within $2^{i+1}\cdot \beta$ writes. Thus, when the buffers are switched after $2^{i+1}\cdot \beta$ writes, all valid content from the merge buffer of level $i$ will be in level $i+1$.
%\end{proofsketch}

\begin{theorem}
	\label{thm:deamort_security}
	The de-amortized merge protocol (Algorithm~\ref{merge_level_deamortized}) ensures that during the merge all writes to level $i$ are 
	uncorrelated and indistinguishable, independent of the logical block addresses.
\end{theorem}

Proofs in Appendix
%\begin{proofsketch} 
%	It is straightforward to see that while executing Algorithm \ref{merge_level_deamortized} (in Appendix), the only steps visible to the adversary are Step 29, 32 and 41. 
%	
%	Step 29 and 41 combined perform a write to the {\em next} block in sequence to the write buffer in every level of the ORAM, irrespective of the logical block address and content. Also,  invariably the process is continued until $\beta$ blocks have been written {\em sequentially} to the write buffer of every level (Step 18). In case, there are less than $\beta$ real data blocks that can be written {\em fake} blocks are written instead. Semantic security ensures that fake blocks are indistinguishable from real data blocks.
%	
%%	Step 32 re-encrypts the content of the {\em next} block in sequence in the last level buffer. Due to semantic security, re-encryption is indistinguishable from a real write, rendering this step indistinguishable from Step 41.  
%	
%	
%\end{proofsketch}

%

%\smallskip
\paragraph{Write access complexity}
For each invocation of the de-amortized merge algorithm after the in-memory write queue has been 
filled up, one bucket is written to each level. Since the write queue is the same size as the buckets, the overall write access complexity is $\O(\log{N})$. Effectively, the de-amortization converts an $\O(\log N)$ amortized construction with a 
worst case of $\O(\log{N})$ to an $\O(\log{N})$ worst case construction.

%\smallskip
\paragraph{Number of seeks}
Observe that once the write queue is filled up (after $\beta$ new writes), the blocks are flushed to the ORAM top level and the de-amortized merge protocol (Algorithm \ref{merge_level_deamortized}) is executed. The protocol writes a bucket to {\em each} of the $\log{N}$ levels. Writing the bucket to a level requires one seek while filling up the two queues with blocks from the previous level merge buffer requires a seek each. In effect, the write queue flush after $\beta$ writes triggers a reshuffle mechanism which requires $s = 3 \log{N}$ seeks. An additional seek is performed for writing the corresponding B-tree buffer map. In other words, the number of seeks performed per write is:

\begin{center}
	$s = \frac{4 \log{N}}{\beta} $
\end{center}

Thus, similar to Section \ref{seqoram:amortized}, with $\beta = \O(\log{N})$, the number of seeks performed by the {\sys} de-amortized construction is a constant.

\subsection{Efficient Reads}
\label{deamortized:reads}

%Since the de-amortized construction (Section \ref{deamortized:writes}) keeps each generation in a level individually sorted, buffer maps have to be maintained 
%for each generation. Consequently, each level has 4 B-trees -- 2 B-trees per 
%generation in each buffer. Also, while searching for a block in a level, the trees need to be checked in order for the most up-to-date location of the block. In particular, this is based on the how recently a generations has been written by a merge. The order is as 
%follows: i) generation 1 in the write buffer, ii) generation 0 in the write buffer, iii) generation 1 in the merge buffer, 
%and iv) generation 0 in the merge buffer.

The asymptotic read complexity for the de-amortized construction is $\O(\log_{\beta}N \times \log N)$. In contrast, position map-based write only ORAMs \cite{usna_stash_free_oram,hive,datalair} benefit from asymptotically faster reads with $\O(\log N)$ access complexity.
The additional read complexity for hierarchical ORAMs is the result of checking 
up to $\log N$ levels in order to locate a particular block, which in turn entails querying per-level indexing structures.

%Note that there may be duplicates in the merge buffer of level $i$ and the write buffer of level $i+1$ while a merge is ongoing. This however is not a problem since as soon as the required block is found in level $i$, the search will stop. 

%
%This is significantly worse than the $\O{log_{\beta}N}$ 
%read complexity for HIVE-ORAM~\cite{hive}.  The additional read complexity for the construction is the
%result of checking all $logN$ levels in the worst case in order to locate a particular block. In contrast, 
%\cite{hive} stores a position map recursively in smaller ORAMs to indicate the physical location of each logical block. 
%The 
%position map is stored recursively in $\order{logN}$ smaller ORAMs. 

%\begin{figure}
%\centering
% \includegraphics[scale=0.12]{figures/access_map_traversal.eps}
% \vspace{-1.5cm}
% \caption{Access time map (ATM) traversal in the ORAM levels. Map blocks are placed together with the 
% data blocks in the levels. Once the level is determined corresponding to each node along an ATM path, the level map 
% is used to retrieve the node.\label{access_map_traversal}}
%\end{figure}

%\smallskip
\paragraph{Reducing read complexity}
Unfortunately, it is non-trivial to track the precise location of 
each block in {\sys}. This is because a block moves down the levels due to periodic reshuffles {\em even when the block is not specifically updated}. Also, the location of each block in a level 
depends on other blocks present in that level.

{\em To perform reads efficiently in {\sys}, our key idea is to correctly predict the level, the buffer and the generation in which a block resides currently}. Then, only the buffer map for that generation can be queried to determine the actual physical location of the block instead of checking buffer maps at all levels. 
This is possible since ORAM writes 
trigger level reconstructions {\em deterministically} --  each 
flush from the write queue is followed by writing exactly {\em one} bucket at each level. So, 
the number of write queue flushes required before the write buffer of a level is full depends only on the level size. In particular, the level in which a block currently resides can be accurately predicted by 
comparing the value of a {\em global access counter} -- tracking the total number of writes since initialization --  and the last access time for the block (mechanism detailed below). The last access time for each block is the value of the global counter when the block 
was flushed from the write queue. 

Using this mechanism, only {\em one} level needs to be checked for a block read,  
reducing the 
asymptotic read access complexity by a factor of $\O(\log N)$. In fact, the idea can be extended further to also correctly predict the buffer and the generation in the level where a particular block currently resides and reduce the constants further. 

First, we describe the mechanism to predict the level, buffer and generation for a a block and then in Section \ref{deamortized:atm} show how to efficiently store the last access time information.

\paragraph{Identifying Generation}
Consider a block with logical address $x$ that was last accessed when the value of the 
global access counter was $c$. Also, let the current value of the global counter be $g$. 
Thus, after $x$ was written, the write queue was flushed $g-c$ times more. During this time, $x$ moved down the levels due to the merge protocol after every flush. 

%
%This is discussed in detail in the appendix.

Predicting the generation to which $x$ will be written in a level is relatively straightforward. Observe that by construction the ORAM is initialized with empty levels -- the first time generation 0 of level $i$ will be full is when it contains all blocks written as part of the first $2^{i}\beta$ writes. In other words, blocks written during the first $2^{i}\beta$ writes (write queue flushed to disk $2^{i}$ times) will be written together to generation 0 in the write buffer in level $i$ due to the periodic execution of the merge protocol. Similarly, the next $2^{i}\beta$ writes will be to generation 1 in the write buffer in level $i$. {\em In particular, the $k^{th}$ group of $2^{i}\beta$ writes will be to generation $k \bmod 2$ in level $i$ write buffer. }

\begin{obs}[Identifying Generation]
If a block $x$ is written when $g = c$, and  $k = \floor{c/2^{i}}$, 
then currently $x$ resides in generation $j$ in level $i$ write buffer where $j = k \bmod 2$.
\end{obs}

%Given the value of $c$, {\sys} first determines $k = \floor{c/2^{i}}$. Effectively this implies that $x$ was written within the $k^{th}$ group of $2^{i}\beta$ writes to the write queue. Then following the discussion above, $k$ will be written to generation $gen_{i} = k \bmod 2$ in level $i$ write buffer.

\paragraph{Identifying Level}
To determine the level in which $x$ currently resides, we specifically track the number of write queue flushes that $x$ spends in  level $i$. 
Based on this, Algorithm \ref{det_level} (in Appendix) determines the level by calculating the cumulative time $x$ spent in all levels $j < i$, for each level $i$ (Lines 5 - 10) and comparing with the total number of write queue flushes that have taken place since $x$ was written (Line 4).

\paragraph{Identifying Buffer}
To identify the correct buffer in which $x$ currently resides in level $i$, we need to determine the number of writes that have taken place since $x$ was written to level $i$. 
% If $x$ was written initially to generation 1 in level$i$, then $x$ will be in generation 1 of the merge buffer of level $i$.
%
% In case, $x$ was written to generation 0 in level $i$, then $x$ will be currently in the merge buffer if $2^{i}$ write queue flushes have  been performed since $x$ was written. The $2^{i}$ write queue flushes correspond to $2^{i}$ buckets in generation 1 of level $i$ write buffer being full. 
%
Specifically, the total number of write queue flushes performed after $x$ was written to level $i$:

\begin{equation}
w = g -c - \sum\limits_{j=1}^{i-1}\mathsf{flush_j}
\end{equation}

Here $\mathsf{flush_j}                                                                                                                                                                                                       $ is the number of write queue flushes $x$ spend in level $j$ as determined by Algorithm \ref{det_level}. The difference in the number of write queue flushes $x$ cumulatively spent in levels 0 to $i-1$ with the total number of write queue flushes that have been performed since $x$ was written determines the number of write queue flushes performed after $x$ was written to level $i$.
  
Thus, in case $x$ is written to generation 0 in level $i$ and $w > 2^{i}$, then $x$ is currently in generation 1 in the merge buffer of level $i$.

% A final caveat to consider here is that once $x$ is in the merge buffer of level $i-1$, the bucket it will 
% be written to in the write buffer in level $i$ will depend on the result of the merge. Since, one bucket 
% is written at a time, it is thus not possible to predict at a given time before a merge finishes whether 
% $x$ has already been written to level $i$ or is still in the merge buffer of level $i-i$. Fortunately, 
% the merge buffer of level $i-i$ is discarded/overwritten only after the merge has completed. Therefore, 
% $x$ can be found in the merge buffer of level $i-1$ before the merge completes

%Now, when writes to generation 1 in the write buffer of level $i$ starts, the 
%buckets in the merge buffer for level $i-1$ is discarded/overwritten. The buckets in generation 0  
%of the write buffer of level $i$ now contains all buckets that were part of the previous merge buffer 
%in level $i-1$. To determine if generation 0 buckets are full in level $i$, it is enough to check if $b + r > 2^{i}$. 
%If true, $x$ can be found in generation 0 of the write buffer. Otherwise, $x$ is still part of the 
%merge buffer in level $i-1$.
%
%To summarize, using the mechanism described above, {\sys} can correctly predict the level number, 
%the buffer and the generation in which a block $x$ currently resides. This allows {\sys} to check only 
%one level 

\begin{figure}
	\centering
	\includegraphics[scale=0.17]{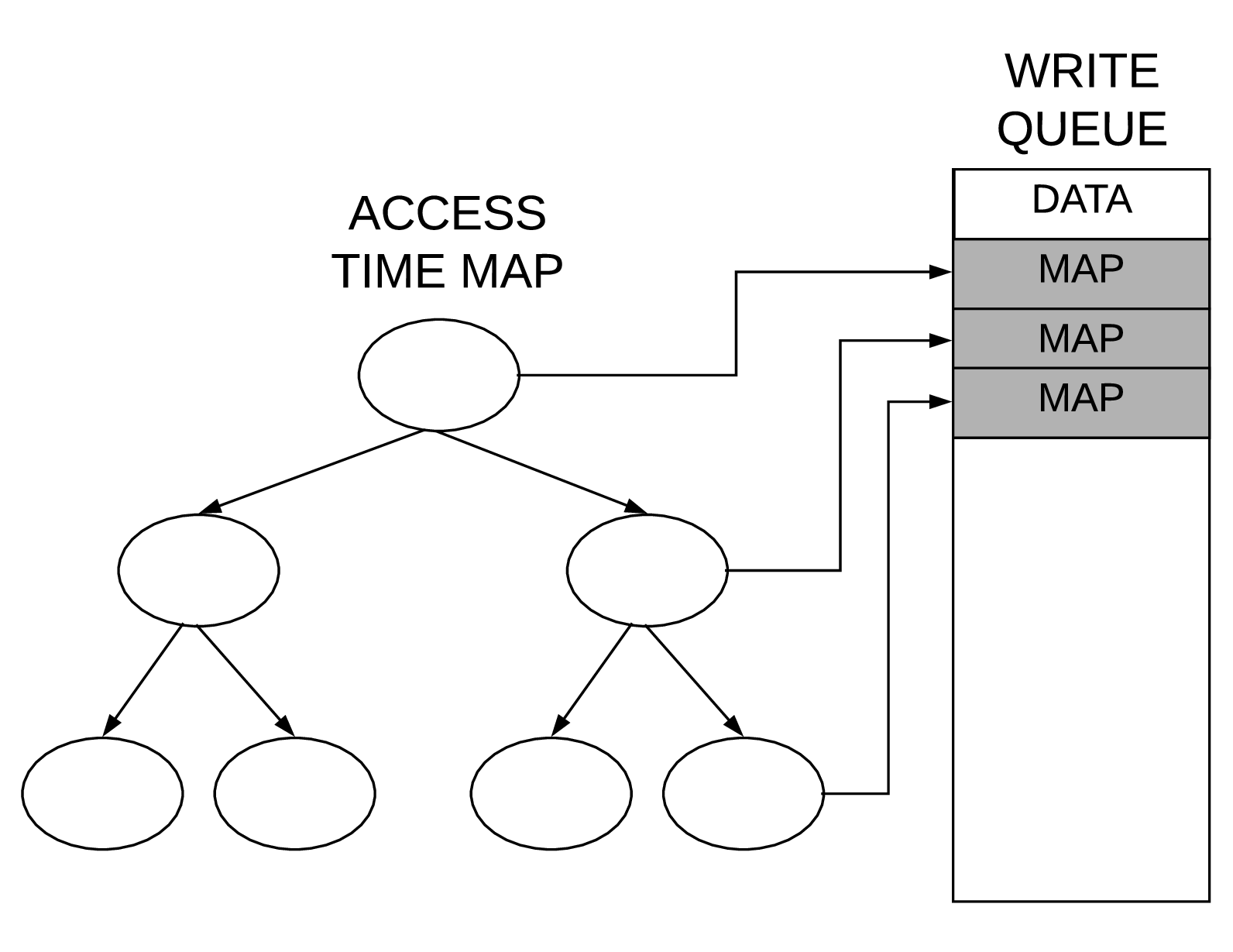}
	\vspace{-0.5cm}
	\caption{\small Writing a path of the ATM in the write queue after updating the path corresponding 
		to an updated data block.\label{atm_to_disk}}
\end{figure}
\noindent
\subsection{Access Time Map}
\label{deamortized:atm}
To use the mechanism described above, we need to track the last time a particular block was accessed. 
For this, {\sys} stores an {\em oblivious data structure} (ODS), named the {\em access time map} (ATM) within the same ORAM with the data. {\em In structure, the ATM is effectively a B+ tree but unlike a standard B+ tree where each node stores pointers to its children nodes, each node in the ATM stores an access time value for its children.} The ATM is traversed from the root to the leaf by determining the location of each child node on the path based on its last access counter value as described above. 

\paragraph{ATM Design}
Each node of the ATM is assigned a logical address within the same address space as the data blocks. 
In particular, each leaf node of the ATM stores a tuple $\mathsf{\<laddr, last\_access\_ctr\>}$  where $\mathsf{laddr}$ 
is the logical address of a block and $\mathsf{last\_access\_ctr}$ is the value of the global access counter 
when the block was last written to the write queue. Recall that the global access counter tracks 
the number of times the write queue has been flushed since the ORAM initialization. Each leaf node 
is stored in one disk block. The number of entries that can fit in a disk block depends on the size of the tuple. 
Assuming 64 bit logical addresses and last access counter values, the number of entries 
in a block can also be fixed as $\beta$ (as defined before). Consequently, 
the height of the tree is $\log_{\beta}N$ with a fanout of $\beta$.

%{\em access time map} (ATM) for efficiently 
%tracking the last access time of a block. The ATM is  stored as a B+ tree (not to be confused 
%with the per level map B+ tree) within the same address space as the data blocks.

The leaf nodes themselves are ordered from left to right 
on the basis of the logical address -- the leftmost leaf node has entries for logical block 
addresses 1 to $\beta$ while the rightmost leaf node has entries for addresses $N - \beta$ to 
$N$. Thus, it is straightforward to determine the path in the tree corresponding to an entry for a
particular block address since the logical block address uniquely defines the corresponding path in the ATM.

Each internal node contains a tuple that keeps track of the 
$\mathsf{\<laddr,last\_access\_ctr\>}$ values of its 
children nodes. The root of the ATM is stored in-memory and allows traversing a path of the ATM by determining locations of the nodes in the ORAM based on the last access counter values.

\paragraph{Querying the ATM}
%
%(Algorithm \ref{read_algo})
To read block $b$ (Algorithm \ref{rw_algo} in Appendix), first the ATM path corresponding to the leaf containing the entry for $b$ is traversed to determine the last access counter value of the block for $b$ (Lines 1 - 9). This determines the level, buffer and generation where the block currently resides in the ORAM. Then the buffer map of the corresponding level, buffer and generation is queried for the location of $b$ and the block is read from there (Line 11 - 13).

The height of the ATM is $\O(\log_{\beta}N)$ and the height of the buffer map is bounded by $\O(\log_{\beta}N)$. Therefore, the overall read complexity is $\O(\log^{2}_{\beta}N)$ -- the ATM reduces the read complexity 
of the de-amortized construction from $\O(\log_{\beta}N \times \log N)$ to $\O(\log^{2}_{\beta}N)$ with $\beta >> 2$.

%\smallskip
\paragraph{Updating the ATM}
%
%(Algorithm \ref{write_algo})
If a data block is to be written/updated (Algorithm \ref{rw_algo} in Appendix), it is first written to the write queue (line 1). This is followed by updating the last accesses counter value for that block in the ATM. Specifically, the leaf node on the ATM path containing an entry for the block is first updated with the new access counter value for the block (Line 4). Then, the path is updated with the new access counter value for the children nodes up to the root (Line 5 - 7), and the update nodes are added to the write queue (line 6).  At this stage, if the write queue is full, it is flushed to the top level and the de-amortized merge protocol (Algorithm \ref{merge_level_deamortized}) is executed (Lines 8 - 10).

%Figure~\ref{atm_to_disk} illustrates how the ATM 
%nodes are written after the updated data block in the write queue. 

\begin{figure}
	\centering
	\includegraphics[scale=0.17]{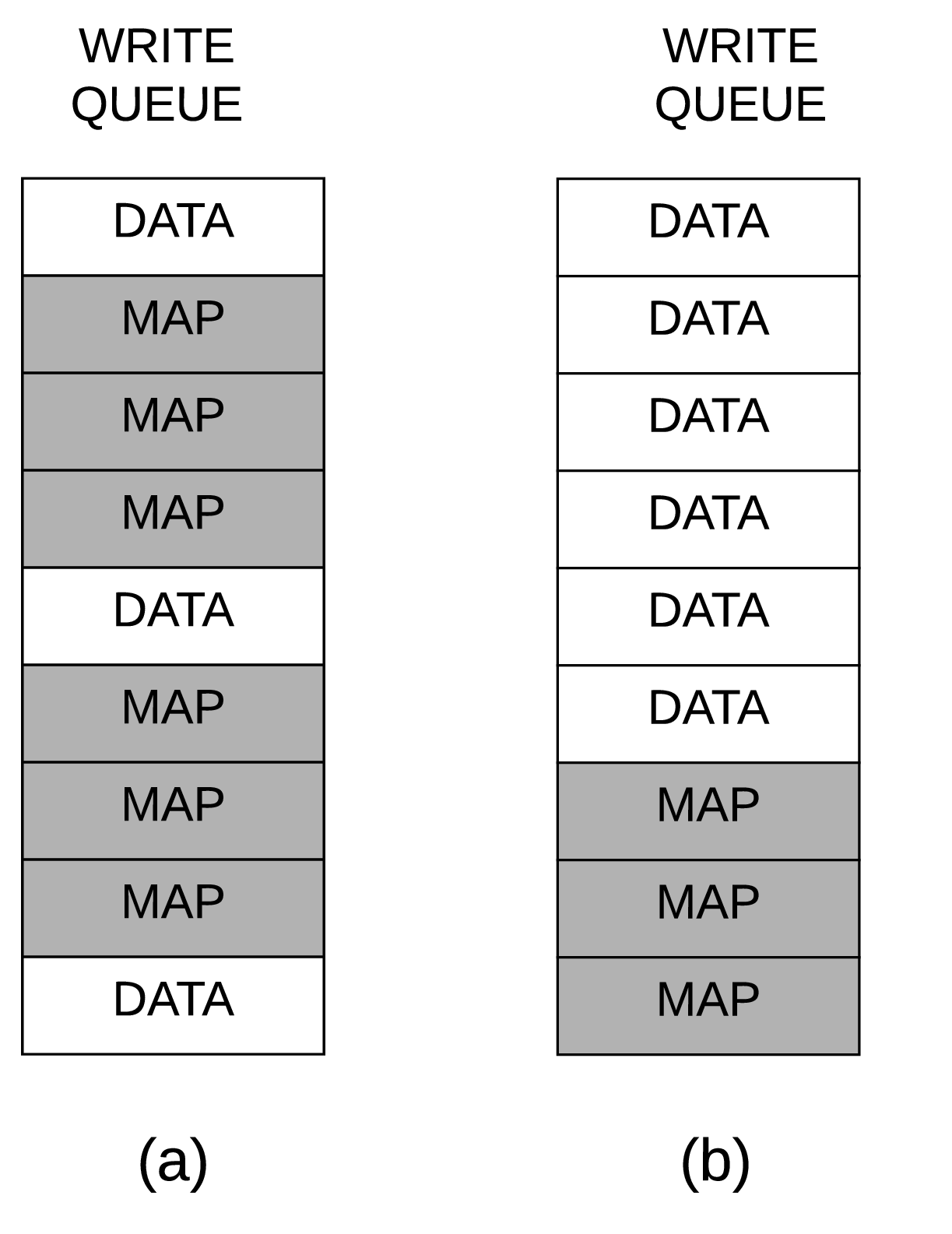}
	\vspace{-0.4cm}
	\caption{\small (a) The state of the write queue when random data blocks are written and the 
		corresponding ATM paths do not intersect. (b) State of the write queue when logically sequential 
		data blocks are written. Only one path common to all the data blocks needs to the be updated for the ATM.
		\label{level_map_tree}}
\end{figure}

Observe that each write of a data block is followed by writing the corresponding path in the ATM
that it needs to update. Thus, the actual number of data blocks that can be written to the write queue before a flush (and consequently the overall write throughput) reduces by a factor equal to the height of the ATM (Figure~\ref{level_map_tree}).

%\subsection{Optimization Knobs}

\paragraph{Sequential reads}
Exploiting sequentiality in logical blocks written together, 
allows optimizing the throughput by writing nodes common to the ATM 
path corresponding to multiple writes only once. For example, in Figure~\ref{level_map_tree}, 
if all the data blocks in the write queue have entries within the same leaf node of the 
ATM, then the updated nodes on the corresponding path can be written only once after all the data writes 
have been completed. This reduces the overhead of writing the same path multiple times. Since, 
the height of the tree, $\log_{\beta}N$ is small (4 for a 1TB database with $\beta = 256$), writing 4 map blocks 
instead of data blocks in a write queue of size $\beta = 256$ leads to minimal reduction in overall throughput. 
Thus, sequential writes are faster than random writes in {\sys}.

%\smallskip
\paragraph{Effect of Caching}
Caching can dramatically improve read throughput by avoiding seeks in 
between sequential reads. In this case, using a cache of $\O(\log_{\beta}N)$ blocks for storing a path 
of the ATM allows optimizing the number of nodes that need to be accessed for the next read. 
If the next read is sequential and has the logical block address within the leaf node in the cache, 
the ATM traversal to locate the level for the block can be completely avoided. 

%In fact, with 
%purely sequential access, this brings down the overall read complexity to $\order{log_{b}N}$ (to access the 
%B+ tree in the corresponding level) since 
%an $\order{log_{b}N}$ sized ATM path is required to be read only after every $\beta$ accesses (the size of the 
%leaf node in the cache) with $\beta >> log_{\beta}N$.

%\smallskip
\paragraph{Map Caches}
To further optimize sequential reads, recently read leaf nodes from the buffer maps can be cached in memory. Since the leaf nodes contain entries sorted on logical addresses, blocks with sequentially increasing logical addresses will have entries within the same leaf node. Thus, for sequential reads, if an entry for a block is found in the ATM path in memory, then instead of querying the buffer map at the required level (and incurring an overhead of $\O(\log_{\beta}N)$), the entry for that particular block will be found in the cached leaf node for that buffer map. 
%In fact proper caching and replacement will ensure that the physical location of a block can always be entirely determined from the contents of the map caches and the cached ATM path in memory. 

\begin{theorem}
 The deamortized {\sys} construction provides write access privacy (Definition \ref{def:privacy}).
\end{theorem}

Proof in Appendix.
%\begin{proofsketch}
%%  
%The de-amortized construction makes two changes to the amortized
%construction: the level reshuffles are performed using the de-amortized merge protocol (Algorithm \ref{merge_level_deamortized}), and the reads are optimized by using the ATM to precisely locate blocks.
%
%Consider two equal-length write access patterns $\vec{A} = {w_1, w_2, \ldots w_i}$ and $\vec{B} = {x_1, x_2, \ldots 	x_i}$. First, note that in the de-amortized construction (similar to the amortized construction) when either of $\vec{A}$ or $\vec{B}$ is executed, the $i$ blocks are added to the write queue irrespective the logical addresses.  Once the write queue is full, its contents are written the top level write buffer, encrypted with semantic security. 
%
%Further, the de-amortized level reshuffling protocol ensures the same security guarantees as the amortized protocol -- while reshuffling level $j < \log N$,  the writes to the disk are uncorrelated to each other and independent of the block addresses (Theorem \ref{thm:deamort_security}). 
%
%Therefore, by simply observing the writes to the top level and the writes due to the level reshuffles, an  adversary can only do negligibly better than purely guessing whether $A$ or $B$ was executed.
%  
%Second, since reads are not observable, using the ATM during reads to
%reduce the number of levels searched for a block does not provide any
%additional information to the adversary. Finally, the ATM blocks are
%written to the write queue indistinguishably from data blocks due to
%semantic security of the encryption.
%%
%\end{proofsketch}

\section{Evaluation}

%\begin{figure*}[t!]	
%\begin{center}
%\subfigure[Sequential read throughput (in MB/s). The x-axis represents size of individual 
%sequential reads (in MB) peformed over a 12GB file. The y-axis is the throughput recorded in logarithmic scale.]{\label{seq_read} 
%\includegraphics[scale=0.60]{figures/plots/sequential_read.eps}}
%\subfigure[Sequential read throughput (in MB/s). The x-axis represents size of individual 
%sequential writes (in MB) peformed over a 12GB file. The y-axis is the throughput recorded in logarithmic scale.]{\label{seq_write} 
%\includegraphics[scale=0.60]{figures/plots/sequential_write.eps}} 
%%\subfigure[Random Read on HDD and Combo]{\label{rand_read} 
%\vspace{-0.2cm}
%\caption{Sequential read and write throughputs for different I/O sizes. Throughputs 
%are in logarithmic scale while I/O size is in MB. Higher is better. \label{seq_log}}
%\end{center}
%\end{figure*}
%
{\sys} has been implemented as a kernel device mapper as well as a virtual
block device using the Block Device in User Space (BUSE) \cite{buse_proj}
framework for a fair comparison with all related work.

\begin{figure*}[h!]
	\subfigure[\small Sequential Access Throughput]{\label{fig_seq}
\begin{tikzpicture}[scale=0.5]
\begin{axis}[
symbolic x coords={SeqRead,SeqWrite},
xtick=data,
enlarge x limits=0.45,
ylabel=Population,
legend style={at={(0.5,-0.15)},
	anchor=north,legend columns=2},
ybar=5pt,% configures `bar shift'
ymin=0,
bar width=11pt,
nodes near coords
]
\addplot 
	coordinates {(SeqRead,59.4) (SeqWrite,32.2)};

\addplot 
coordinates {(SeqRead,18) (SeqWrite,16.1)};

\addplot 
coordinates {(SeqRead,30) (SeqWrite,4.8)};

\addplot 
coordinates {(SeqRead,48.6) (SeqWrite,6.1)};

\legend{Baseline,DetWoORAM,SqORAM (on-disk ATM),SqORAM (in-memory ATM)}
\end{axis}
\end{tikzpicture}
}
	\subfigure[\small Random Access Throughput]{\label{fig_rand}
	\begin{tikzpicture}[scale=0.5]
	\begin{axis}[
	symbolic x coords={RandRead,RandWrite},
	xtick=data,
	enlarge x limits=0.45,
	ylabel=Population,
	ymin=0,
	legend style={at={(0.5,-0.15)},
		anchor=north,legend columns=2},
	ybar=6pt,% configures `bar shift'
	bar width=11pt,
	nodes near coords
	]
	\addplot 
	coordinates {(RandRead,1.7) (RandWrite,4.6)};
	
	\addplot 
	coordinates {(RandRead,0.4) (RandWrite,12.0)};
	
	\addplot 
	coordinates {(RandRead,0.12) (RandWrite,1.2)};
	
	\addplot 
	coordinates {(RandRead,0.7) (RandWrite,5.0)};
		
\legend{Baseline,DetWoORAM,SqORAM (on-disk ATM),SqORAM (in-memory ATM)}
	\end{axis}
	\end{tikzpicture}
}
	\subfigure[\small Workload Benchmarks]{\label{fig_bench}
	\begin{tikzpicture}[scale=0.5]
	\begin{axis}[
	            x tick label style  = {text width=2cm,align=center},
		symbolic x coords={R=30\% W=70\%,R=60\% W=40\%,R=70\% W=30\%},
	xtick=data,
	ylabel=Population,
	enlarge x limits=0.25,
	legend style={at={(0.5,-0.25)},
		anchor=north,legend columns=-1},
	ybar=6pt,% configures `bar shift'
	bar width=11pt,
	nodes near coords
	]
	\addplot 
	coordinates {(R=30\% W=70\%,48.7) (R=60\% W=40\%,48.5) (R=70\% W=30\%,48)};
	
	\addplot 
	coordinates {(R=30\% W=70\%,20) (R=60\% W=40\%,21) (R=70\% W=30\%,19.1)};
	
	\addplot 
	coordinates {(R=30\% W=70\%,15) (R=60\% W=40\%,23.5) (R=70\% W=30\%,31.3)};
	
	\legend{Baseline,DetWoORAM,SqORAM}
	\end{axis}
	\end{tikzpicture}
}
\vspace{-0.5cm}
\caption{\small Throughput comparison in MB/s (higher is better). The baseline is a block device implemented in BUSE that translates FS requests to block requests.
	(a) When the ATM is stored in memory, {\sys} can achieve almost raw-disk sequential read throughputs and slightly better performance for random reads compared to DetWoORAM. 
	When the ATM is on disk, {\sys} sequential read throughput is higher (almost 2x)
	than DetWoORAM sequential read throughput. (b) SqORAM outperforms DetWoORAM for random reads. (c) Throughput comparison in MB/s (higher is better) for different read/write distributions. With 70\% reads and 30\% writes , {\sys} is 1.6x faster than {\usnaoram} \cite{usna_stash_free_oram}. DetWoORAM performs generally better for write-intensive workloads. \label{micro_bench}}
\end{figure*}

\subsection{Kernel-space Implementation} 
%
% {\color {blue}
% 
{\sys} has been implemented as a kernel device mapper and benchmarked in
comparison with the kernel implementations of HIVE \cite{hive} and
DataLair \cite{datalair}
%  \footnote{{\color {blue} When this paper was
% written, HIVE-ORAM \cite{hive} was the state-of-the-art write-only ORAM.  We
% include comparison with DataLair \cite{datalair} based on the revised
% construction \cite{datalair_fix} using code provided by the authors.}}
% 
% }

The cipher used for encryption is AES-CTR (256 bit) with individual
per-block random IVs.  IVs are stored in a pre-allocated location on disk. 
Underlying hardware blocks are 512 bytes each and 8 adjacent hardware blocks
constitute a ``physical block'' (4KB in size).

%\smallskip
\paragraph{Setup}
Benchmarks were conducted on Linux boxes with Intel Core i7-3520M processors
running at 2.90GHz and 4GB+ of DDR3 DRAM.  The storage device of choice was
a 1TB IBM 43W7622 SATA HDD running at 7200 RPM.  The average seek time and
rotational latency of the disk is 9ms and 4.17ms respectively.  The data
transfer rate is 300MB/s.

{\sys} was built on a 256GB physical partition.  Benchmarks were performed
using FileBench version 1.4.9.1 on Ubuntu 14.04 LTS, kernel version 3.13.6. 
Results for HIVE~\cite{hive} were collected by compiling the open
source project~\cite{hive_proj}. We thank the authors of DataLair \cite{datalair} for providing 
their implementation. All tests were run multiple times with same parameters and results were collected with a
95\% confidence interval.

%\smallskip
\paragraph{Results}
Tests were performed using the sequential and random read/write workload
personalities of FileBench.  Sequential accesses were measured over a 8GB
file by performing individual 1MB sequential IOs.  Using a file size twice
the size of the available DRAM (4GB here) eliminates caching effects.  For random
reads/writes, individual I/O sizes were reduced to 4KB. 
Table~\ref{throughput_kernel} compares the sequential and random read/write
throughputs for {\sys} with HIVE, DataLair and dm-crypt, a Linux device mapper for disk encryption.

{\sys} is almost 105x faster than HIVE~\cite{hive} for sequential reads
and 100x faster for sequential writes.  Random write performance for {\sys}
and HIVE~\cite{hive} are comparable while HIVE~\cite{hive} peforms
better for random reads as it features a read complexity of
$\O(\log_{\beta}N)$ compared to $\O(\log_{\beta}^{2}N)$ for {\sys}. 

\begin{table*}
	\begin{center}
		\tabcolsep=0.11cm
		\begin{tabular}{ | l | l | l | l |l| }
			\hline
			Access & dm-crypt & {\sys} & HIVE \cite{hive} & DataLair~\cite{datalair} \\ 
			\hline
			Sequential Read  & 91  & 21 & 0.135 & 0.200\\
			Sequential Write & 88 & 1.5 & 0.016 & 0.110 \\
			Random Read & 5.0  & 0.055 & 0.120 & 0.105\\
			Random Write & 4.3 & 1.0 & 0.014 & 0.2\\
			\hline
		\end{tabular}
		\vspace{3pt}
		\caption{\small Throughput comparison in MB/s (higher is better). {\sys} features a 150x speedup over HIVE~\cite{hive} and DataLair \cite{datalair}
			for sequential reads and a 100x speedup for sequential writes. {\sys} random read performance is
			comparable to HIVE~\cite{hive}.
			\label{throughput_kernel}}
	\end{center}
\end{table*}

\subsection{Userland Implementation}
To compare with the user space implementation of DetWoORAM
\cite{detwooram_proj},  {\sys} has also been implemented as a virtual block
device using the Block Device in User Space (BUSE) \cite{buse_proj}
framework.  
% BUSE mounts a partition with a driver written
% in user space as a block device through the Linux (network block device) NBD
% framework.  
% 
The baseline is a block device which translates file system
requests into block requests to the underlying device. 
It is necessary to build a baseline from the ground up since, as also noted in
\cite{usna_stash_free_oram}, simply using a loopback device with BUSE as the
baseline unfairly overestimates performance by directly offsetting
arbitrary lengths within the partition without any address translation to
blocks.

Tests were run on a 40 GB ORAM (similar parameters as
used in \cite{usna_stash_free_oram}).  Each ORAM volume was mounted using an
ext4 file system.  DetWoORAM was setup with the holding area equal to thrice
the size of the main area.  Filebench results are presented in Figure
\ref{micro_bench}.
\begin{table*}
	\begin{center}
		\tabcolsep=0.11cm
		\begin{tabular}{ | l | l | l | l |}
			\hline
			ORAM & Before random write & After random write\\ 
			\hline
			Baseline  & 61  & 60.7 \\
			{\sys} & 47.2 & 42  \\
			DetWoORAM \cite{usna_stash_free_oram} & 18.7 & 10.2\\
			
			\hline
		\end{tabular}
		%\vspace{-2pt}
		\caption{\small Throughput comparison. (MB/s, higher is better). 
			A series of block-sized random updates to a large file, sequentially-written
			on disk by DetWoORAM, results in a drop in the sequential read throughput while reading the file subsequently.  The drop
			in throughput for the baseline and {\sys} are not significant.
			\label{throughput_rand_read}}
		
	\end{center}
\end{table*}
 
\paragraph{Micro-Benchmark Results}
 DetWoORAM performs  better for
writes since it performs two physical writes for each logical write in
contrast to the $\log N$ worst case writes of {\sys}.  Both logically random and
sequential writes in DetWoORAM result in the same physical writes by construction.

Being optimized for reads, {\sys} outperforms {\usnaoram} for sequential reads. The advantages of maintaining data locality can be clearly observed in the
sequential read throughput, where the overhead compared to the baseline is
less than 2x. In fact, for memory-rich systems if the ATM is stored
in-memory, {\sys} can achieve sequential read throughputs close to the
baseline.  For a 40GB partition, the ATM requires 128MB of memory
considering 64 bit access time counters. Note that trivially storing in-memory maps for DetWoORAM will not result in similar gains as logically sequential data is not maintained close on disk throughout its lifetime due to frequent updates.

% {\color{blue}
Interestingly, for random writes both {\usnaoram} and {\sys} outperform the baseline. This is because, for the baseline, the logical address of a block determines the physical address where the data is written on disk. Thus, writing to random logical addresses, incurs a large amount of disk seeks. For both DetWoORAM and SqORAM, the physical addresses for performing writes are not correlated to the logical addresses, and {\em all} writes are performed while preserving locality of access. 
%}

\paragraph{Effect of Workload Distribution}
To further investigate the effects of the read/write distribution in workloads on the performance of {\sys}, we evaluated three FileBench workloads with {\sys} and {\usnaoram} \cite{usna_stash_free_oram} with different read-write distributions. For the read-intensive workloads (> 50\% reads), {\sys} is almost 1.7x faster than {\usnaoram} and only 1.5x slower than the baseline. Note that typical file system workloads are read-intensive \cite{survey_network_FS,survey_typical_FS}. For the write-intensive workload (e.g., for online backup services), {\usnaoram} is generally faster than {\sys}.

%\begin{enumerate}
%	\item A write-intensive FTP file server workload which includes around 60\% writes and 40\% reads.
%	\item A read-intensive mail server (varmail) workload which includes around 70\% reads and 30\% writes, similar to a typical file system workload \cite{web_server_read_write_dist}.
%\end{enumerate}

\paragraph{File System Aging \& Effect of Micro-Writes}
As noted before, the physical layout of DetWoORAM is similar to a log-structured file system. 
It is well known that log structured filesystems perform poorly for reads --
performance degrades over time as the file system {\em ages} -- as an
increasing number of smaller random writes (updates) are performed across large
sequential files \cite{betrfs}.  This results in a file's blocks being
scattered, making sequential reads more expensive.  Standard micro
benchmarks do not capture this behavior since in most cases, sequential reads are
performed on a sequentially written file without interleaving random writes.

To understand the effects of {\em micro-writes}, we use the sequential read
after random write benchmark \cite{betrfs}.  The test
writes a large file (1GB) sequentially, followed by a sequential read. 
Then, a fixed number (around 100MB) of random {\em block-sized} random
writes are performed on the file.  This is followed by sequentially reading
the file again and comparing the read throughput with the earlier reported
value.  Results are tabulated in Table \ref{throughput_rand_read}.

The sequential read throughputs for the baseline and {\sys} remain largely
unaffected due to data locality -- logically-close blocks remain close 
on disk.  DetWoORAM throughput however drops since blocks are increasingly scattered
and additional seeks are required.

%We note that with faster refreshing of the main area with blocks from the holding area, the drop in throughput can be reduced. 

%thus providing a complimentary solution to DetWoORAM.  

%\smallskip
\paragraph{Memory Footprint}
{\sys} does not require more than 30MB (mostly for caches and queues), even for 1TB+ ORAMs. The full access time map for a 40GB ORAM takes up less than 128MB, and if stored in memory, the total memory footprint for {\sys} is upper-bound by 200MB. In that case, periodically syncing the ATM with an on-disk copy can ensure crash consistency.

%{\sys} employs caches and stores (a partial) access time map to improve
%sequential performance as shown above.  \todo{This does not pose consistency
%issues -- (discuss why) and is acceptable in most scenarios} -- e.g., . 
%Alternately, the ATM can be stored on disk.  In that case, .
% 
% }

\section{Conclusion}

{\sys} is a write-only ORAM that achieves write access privacy while
preserving locality of access for both reads and writes. {\sys}  maintains an increased level of data locality over time, thus significantly increasing throughput for sequential reads.  {\sys} is 60-100\% faster than the existing state-of-the-art for typical file system workloads.

\begin{scriptsize}
\bibliographystyle{abbrv}
\bibliography{Bib/references}

% Generated by IEEEtranSN.bst, version: 1.14 (2015/08/26)
\begin{thebibliography}{21}
\providecommand{\natexlab}[1]{#1}
\providecommand{\url}[1]{#1}
\csname url@samestyle\endcsname
\providecommand{\newblock}{\relax}
\providecommand{\bibinfo}[2]{#2}
\providecommand{\BIBentrySTDinterwordspacing}{\spaceskip=0pt\relax}
\providecommand{\BIBentryALTinterwordstretchfactor}{4}
\providecommand{\BIBentryALTinterwordspacing}{\spaceskip=\fontdimen2\font plus
\BIBentryALTinterwordstretchfactor\fontdimen3\font minus
  \fontdimen4\font\relax}
\providecommand{\BIBforeignlanguage}[2]{{%
\expandafter\ifx\csname l@#1\endcsname\relax
\typeout{** WARNING: IEEEtranSN.bst: No hyphenation pattern has been}%
\typeout{** loaded for the language `#1'. Using the pattern for}%
\typeout{** the default language instead.}%
\else
\language=\csname l@#1\endcsname
\fi
#2}}
\providecommand{\BIBdecl}{\relax}
\BIBdecl

\bibitem[Aviv et~al.(2017)Aviv, Choi, Mayberry, and Roche]{oblivisync}
A.~J. Aviv, S.~G. Choi, T.~Mayberry, and D.~S. Roche, ``Oblivisync: Practical
  oblivious file backup and synchronization,'' in \emph{24th Annual Network and
  Distributed System Security Symposium, {NDSS} 2017, San Diego, California,
  USA, February 26 - March 1, 2017}, 2017.

\bibitem[Blass et~al.()Blass, Mayberry, Noubir, and Onarlioglu]{hive_proj}
E.-O. Blass, T.~Mayberry, G.~Noubir, and K.~Onarlioglu, ``Hive,''
  \url{"http://www.onarlioglu.com/hive" }.

\bibitem[Blass et~al.(2014)Blass, Mayberry, Noubir, and Onarlioglu]{hive}
E.~Blass, T.~Mayberry, G.~Noubir, and K.~Onarlioglu, ``Toward robust hidden
  volumes using write-only oblivious {RAM},'' in \emph{Proceedings of the 2014
  {ACM} {SIGSAC} Conference on Computer and Communications Security,
  Scottsdale, AZ, USA, November 3-7, 2014}, 2014, pp. 203--214.

\bibitem[Chakraborti et~al.(2017)Chakraborti, Chen, and Sion]{datalair}
\BIBentryALTinterwordspacing
A.~Chakraborti, C.~Chen, and R.~Sion, ``Datalair: Efficient block storage with
  plausible deniability against multi-snapshot adversaries,'' \emph{PoPETs},
  vol. 2017, no.~3, p. 179, 2017. [Online]. Available:
  \url{https://doi.org/10.1515/popets-2017-0035}
\BIBentrySTDinterwordspacing

\bibitem[Chen et~al.(2019)Chen, Chakraborti, and Sion]{chenPets19}
C.~Chen, A.~Chakraborti, and R.~Sion, ``{PD-DM:} an efficient
  locality-preserving block device mapper with plausible deniability,''
  \emph{PoPETs}, vol. 2019, no.~1, pp. 153--171, 2019.

\bibitem[Cozzette()]{buse_proj}
A.~Cozzette, ``Block device in user space (buse),''
  \url{"https://github.com/acozzette" }.

\bibitem[Goldreich and Ostrovsky(1996)]{goldreich}
O.~Goldreich and R.~Ostrovsky, ``Software protection and simulation on
  oblivious rams,'' \emph{Journal of the ACM}, vol.~43, pp. 431--473, 1996.

\bibitem[Goodrich et~al.(2011)Goodrich, Mitzenmacher, Ohrimenko, and
  Tamassia]{goodrich_deamortized}
\BIBentryALTinterwordspacing
M.~T. Goodrich, M.~Mitzenmacher, O.~Ohrimenko, and R.~Tamassia, ``Oblivious ram
  simulation with efficient worst-case access overhead,'' in \emph{Proceedings
  of the 3rd ACM Workshop on Cloud Computing Security Workshop}, ser. CCSW
  '11.\hskip 1em plus 0.5em minus 0.4em\relax New York, NY, USA: ACM, 2011, pp.
  95--100. [Online]. Available:
  \url{http://doi.acm.org/10.1145/2046660.2046680}
\BIBentrySTDinterwordspacing

\bibitem[Haider and van Dijk(2016)]{flatoram}
S.~K. Haider and M.~van Dijk, ``Flat {ORAM:} {A} simplified write-only
  oblivious {RAM} construction for secure processor architectures,''
  \emph{CoRR}, vol. abs/1611.01571, 2016.

\bibitem[Islam et~al.(2012)Islam, Kuzu, and Kantarcioglu]{accesspatternleak}
\BIBentryALTinterwordspacing
M.~S. Islam, M.~Kuzu, and M.~Kantarcioglu, ``Access pattern disclosure on
  searchable encryption: Ramification, attack and mitigation,'' in \emph{19th
  Annual Network and Distributed System Security Symposium, {NDSS} 2012, San
  Diego, California, USA, February 5-8, 2012}, 2012. [Online]. Available:
  \url{https://www.ndss-symposium.org/ndss2012/access-pattern-disclosure-searchable-encryption-ramification-attack-and-mitigation}
\BIBentrySTDinterwordspacing

\bibitem[Jannen et~al.(2015)Jannen, Yuan, Zhan, Akshintala, Esmet, Jiao,
  Mittal, Pandey, Reddy, Walsh, Bender, Farach-Colton, Johnson, Kuszmaul, and
  Porter]{betrfs}
\BIBentryALTinterwordspacing
W.~Jannen, J.~Yuan, Y.~Zhan, A.~Akshintala, J.~Esmet, Y.~Jiao, A.~Mittal,
  P.~Pandey, P.~Reddy, L.~Walsh, M.~Bender, M.~Farach-Colton, R.~Johnson, B.~C.
  Kuszmaul, and D.~E. Porter, ``Betrfs: A right-optimized write-optimized file
  system,'' in \emph{Proceedings of the 13th USENIX Conference on File and
  Storage Technologies}, ser. FAST'15.\hskip 1em plus 0.5em minus 0.4em\relax
  Berkeley, CA, USA: USENIX Association, 2015, pp. 301--315. [Online].
  Available: \url{http://dl.acm.org/citation.cfm?id=2750482.2750505}
\BIBentrySTDinterwordspacing

\bibitem[Kushilevitz et~al.(2012)Kushilevitz, Lu, and
  Ostrovsky]{kushilevitzoram}
E.~Kushilevitz, S.~Lu, and R.~Ostrovsky, ``On the (in) security of hash-based
  oblivious ram and a new balancing scheme,'' in \emph{Proceedings of the
  twenty-third annual ACM-SIAM symposium on Discrete Algorithms}.\hskip 1em
  plus 0.5em minus 0.4em\relax SIAM, 2012, pp. 143--156.

\bibitem[Leung et~al.(2008)Leung, Pasupathy, Goodson, and
  Miller]{survey_network_FS}
\BIBentryALTinterwordspacing
A.~W. Leung, S.~Pasupathy, G.~Goodson, and E.~L. Miller, ``Measurement and
  analysis of large-scale network file system workloads,'' in \emph{USENIX 2008
  Annual Technical Conference}, ser. ATC'08.\hskip 1em plus 0.5em minus
  0.4em\relax Berkeley, CA, USA: USENIX Association, 2008, pp. 213--226.
  [Online]. Available: \url{http://dl.acm.org/citation.cfm?id=1404014.1404030}
\BIBentrySTDinterwordspacing

\bibitem[Li and Datta(2017)]{li_write_only}
\BIBentryALTinterwordspacing
L.~Li and A.~Datta, ``Write-only oblivious ram-based privacy-preserved access
  of outsourced data,'' \emph{Int. J. Inf. Secur.}, vol.~16, no.~1, pp. 23--42,
  Feb. 2017. [Online]. Available:
  \url{https://doi.org/10.1007/s10207-016-0329-x}
\BIBentrySTDinterwordspacing

\bibitem[Peters et~al.(2015)Peters, Gondree, and Peterson]{defy}
T.~Peters, M.~Gondree, and Z.~N.~J. Peterson, ``{DEFY:} {A} deniable, encrypted
  file system for log-structured storage,'' in \emph{22nd Annual Network and
  Distributed System Security Symposium, {NDSS} 2015, San Diego, California,
  USA, February 8-11, 2014}, 2015.

\bibitem[Roche et~al.()Roche, Aviv, Choi, and Mayberry]{detwooram_proj}
D.~S. Roche, A.~J. Aviv, S.~G. Choi, and T.~Mayberry, ``Deterministic
  stash-free write-only oram,'' \url{"https://github.com/dsroche/detworam" }.

\bibitem[Roche et~al.(2017)Roche, Aviv, Choi, and
  Mayberry]{usna_stash_free_oram}
\BIBentryALTinterwordspacing
D.~S. Roche, A.~Aviv, S.~G. Choi, and T.~Mayberry, ``Deterministic, stash-free
  write-only oram,'' in \emph{Proceedings of the 2017 ACM SIGSAC Conference on
  Computer and Communications Security}, ser. CCS '17.\hskip 1em plus 0.5em
  minus 0.4em\relax New York, NY, USA: ACM, 2017, pp. 507--521. [Online].
  Available: \url{http://doi.acm.org/10.1145/3133956.3134051}
\BIBentrySTDinterwordspacing

\bibitem[Roselli et~al.(2000)Roselli, Lorch, and Anderson]{survey_typical_FS}
\BIBentryALTinterwordspacing
D.~Roselli, J.~R. Lorch, and T.~E. Anderson, ``A comparison of file system
  workloads,'' in \emph{Proceedings of the Annual Conference on USENIX Annual
  Technical Conference}, ser. ATEC '00.\hskip 1em plus 0.5em minus 0.4em\relax
  Berkeley, CA, USA: USENIX Association, 2000, pp. 4--4. [Online]. Available:
  \url{http://dl.acm.org/citation.cfm?id=1267724.1267728}
\BIBentrySTDinterwordspacing

\bibitem[Stefanov et~al.(2013)Stefanov, van Dijk, Shi, Fletcher, Ren, Yu, and
  Devadas]{pathoram}
\BIBentryALTinterwordspacing
E.~Stefanov, M.~van Dijk, E.~Shi, C.~Fletcher, L.~Ren, X.~Yu, and S.~Devadas,
  ``Path oram: An extremely simple oblivious ram protocol,'' in
  \emph{Proceedings of the 2013 ACM SIGSAC Conference on Computer \&\#38;
  Communications Security}, ser. CCS '13.\hskip 1em plus 0.5em minus
  0.4em\relax New York, NY, USA: ACM, 2013, pp. 299--310. [Online]. Available:
  \url{http://doi.acm.org/10.1145/2508859.2516660}
\BIBentrySTDinterwordspacing

\bibitem[Williams et~al.(2008)Williams, Sion, and Carbunar]{bforam}
\BIBentryALTinterwordspacing
P.~Williams, R.~Sion, and B.~Carbunar, ``Building castles out of mud: Practical
  access pattern privacy and correctness on untrusted storage,'' in
  \emph{Proceedings of the 15th ACM Conference on Computer and Communications
  Security}, ser. CCS '08.\hskip 1em plus 0.5em minus 0.4em\relax New York, NY,
  USA: ACM, 2008, pp. 139--148. [Online]. Available:
  \url{http://doi.acm.org/10.1145/1455770.1455790}
\BIBentrySTDinterwordspacing

\bibitem[Williams et~al.(2012)Williams, Sion, and Tomescu]{privatefs}
\BIBentryALTinterwordspacing
P.~Williams, R.~Sion, and A.~Tomescu, ``Privatefs: A parallel oblivious file
  system,'' in \emph{Proceedings of the 2012 ACM Conference on Computer and
  Communications Security}, ser. CCS '12.\hskip 1em plus 0.5em minus
  0.4em\relax New York, NY, USA: ACM, 2012, pp. 977--988. [Online]. Available:
  \url{http://doi.acm.org/10.1145/2382196.2382299}
\BIBentrySTDinterwordspacing

\end{thebibliography}
\end{scriptsize}
\section{Appendix}
\begin{algorithm}[th!]
		\caption{$\mathsf{merge(i)}$\label{merge_level}}
	%\begin{algorithmic}[1]
		\scriptsize
		$q_w := \phi$ (queue of size $\beta$)\;
		$q_m:= \phi$ (queue of size $\beta$)\;
		$ctr_{w} := 0$ \;
		$ctr_{m} := 0$ \;
		$ctr_{next} := 0$\;
		\For{x = 1 to $2^{i+1}\beta$}{
		%	\If{$q_0.empty$}				
		\While{ $q_w.notFull$}{
		\If{$ctr_{w} \leq 2^{i}\beta$}{
		Enqueue($q_w$, readFromWriteBuffer($i, ctr_{w}$))\;
		$ctr_{w} = ctr_{w} + 1$\;
		}
		\Else{
		Enqueue($q_w$, fake)\;
		}
		}
		
		%	\EndIf
		%	\If{$q_1.empty$}		
		\While{ $q_m.notFull$}{
		\If{$ctr_{m} \leq 2^{i}\beta$}{
		Enqueue($q_m$, readFromMergeBuffer($i, ctr_{m}$))\;
		$ctr_{m} = ctr_{m} + 1$ \;
		}
		\Else{
		Enqueue($q_m$, fake)\;
		}
		}
		}
		%	\EndIf
		\For{y = 1 to $\beta$}{
		$b_0 = q_{0}.peek$\;
		$b_1 = q_{1}.peek$\;
		%	\If{$i \neq \log{N}$}
		\If {$b_0.add < b_1.add$}{
		$b = \mathsf{Dequeue}(q_{0})$\;
		}
		\ElseIf {$b_0.add = b_1.add$}{
		$b = \mathsf{Dequeue}(q_{0})$\;
		(discard $b_1$ from $q_{1}$)\;
		}
		\Else{
		$b = \mathsf{Dequeue}(q_{1})$\;
		}
		\If{$i \neq \log{N}$}{
		writeNextToWriteBuffer($i+1, ctr_{next}, b$)\;
		}
		\Else{
		writeNextToWriteBuffer($i, ctr_{next}, b$)\;
		}
		$ctr_{next} = ctr_{next} + 1$\;
		}
		%\end{footnotesize}							
	%\end{algorithmic}
\end{algorithm}

\begin{algorithm}[th!]
\caption{$\mathsf{merge\_deamortized}$}\label{merge_level_deamortized}
		%\begin{scriptsize}
		\scriptsize
	      \SetKwFunction{algo}{algo}\SetKwFunction{proc}{proc}
	    \SetKwProg{myproc}{Procedure}{}{}
	    \myproc{$Init(\phi)$}{
		$q_{x0} := \phi$ // Persistent in-memory queue of size $\beta$ assigned for level $x < \log{N}$\;
		$q_{x1} := \phi$ // Persistent in-memory queue of size $\beta$ assigned for level $x < \log{N}$\;
		$ctr_{x0} := 0$ // Persistent in-memory counter assigned for level $x < \log{N}$ \; 
		$ctr_{x1} := 0$ // Persistent in-memory counter assigned for level $x < \log{N}$\; 
		$g := 0 $ // Global Access Counter (number of times write queue has been flushed to the disk)\;
	    }
		\SetKwProg{myproc}{Procedure}{}{}
	    \myproc{$\mathsf{Merge}(\phi)$}{
		\For{x = 1 to $\log{N} - 1$}{
		    \If{$x \neq \log{N} - 1$}{ 
		        $ctr_{next} := (g \bmod 2^{x+1}) \times \beta$\;
		     }
		    \Else{
		        $ctr_{next} := (g \bmod 2^{x}) \times \beta$\;
		    }
		    \While{ $q_{x0}.notFull$}{
		        \If{$ctr_{0} \leq 2^{i}\beta$}{
		            $\mathsf{Enqueue}(q_{x0}, \mathsf{readFromMergeBuffer}(i, ctr_{x0}))$\; 
		            $ctr_{x0} = (ctr_{x0} +1) \bmod 2^{x} $\;
		        }
	            \Else{
		            $\mathsf{Enqueue}(q_{x0}, fake)$\;
	            }
		    }
		    \While{ $q_{x1}.notFull$}{
		        \If{$ctr_{x1} \leq 2^{i}\beta$}{
		            $\mathsf{Enqueue}(q_{x1}, \mathsf{readFromMergeBuffer}(i, ctr_{x1} + 2^{x} \times \beta$))\; 
		            $ctr_{x1} = (ctr_{x1} +1) \bmod 2^{x}$\;
		            }
	    	    \Else{
		            $\mathsf{Enqueue(q_{x1}, fake)}$\;
		        }
		    }
		\For{y = 1 to $\beta$}{
		    $b_0 = q_{x0}.peek$\;
		    $b_1 = q_{x1}.peek$\;
		    \If{$i \neq \log{N} - 1$}{
		        \If {$b_0.add < b_1.add$}{
		            $b = \mathsf{Dequeue}(q_{x0})$\;
		        }
		        \ElseIf {$b_0.add = b_1.add$}{
		            $b = \mathsf{Dequeue}(q_{x0})$\;
		            (discard $b_1$ from $q_{x1}$)\;
		        }
		        \Else{
		            $b = \mathsf{Dequeue}(q_{x1})$\;
		        }
		        writeNextToWriteBuffer($x+1, ctr_{next}, b$)\;
		        }
		    \Else{
		        \If{$b_0.add, b_1.add \neq ctr_{next}$}{
		            Reencrypt($i, ctr_{next}$)\;
	            }
	                
		        \If {$b_0.add < b_1.add$}{
		            $b = \mathsf{Dequeue}(q_{0})$\;
		        }
		        \ElseIf {$b_0.add = b_1.add$}{
		            $b = \mathsf{Dequeue}(q_{0})$\;
	                (discard $b_1$ from $q_{1}$)\;
	            }
		        \Else{
		            $b = \mathsf{Dequeue}(q_{1})$\;
		        }
		        writeToLastLevelBuffer($x+1, ctr_{next}, b$)\;
		    }
        }
            	$ctr_{next} = ctr_{next} + 1$	\;

	}
		}

		%\end{scriptsize}							
\end{algorithm}
\begin{algorithm}[th!]
	\caption{$\mathsf{Determine\_Level(g,c)}$}\label{det_level}
	\scriptsize
		%	\begin{footnotesize}
		$\mathsf{flush_i} = 0$ // Number of subsequent write queue flushes $x$ spent in level $i$ after being flushed to the top level	\;
		$\mathsf{flush_{sum}} = 0$ \; 	
		$i = 0$\;
		\While{$\mathsf{flush_{sum}} \leq g - c$}{
		    \If{$\floor{c/2^{i}}\bmod 2 = 0$}{
		        ($x$ was written to generation 0 in level $i$)\;
		        $\mathsf{flush_i} = 3 \times 2^{i}$\;
		        }
		    \Else{
		        ($x$ was written to generation 1 in level $i$)\;
		        $\mathsf{flush_i} = 2 \times 2^{i}$\;
		    }
		$i = i+1$ \;
		$\mathsf{flush_{sum}} = \mathsf{flush_{sum}} + \mathsf{flush_i}$ \;
		}
		\Return $i-1$\;
	\end{algorithm}
%
% %
\begin{algorithm}[th!]
	\caption{$\mathsf{SqORAM\_Access}$}\label{rw_algo}
	\scriptsize
	    \SetKwFunction{algo}{algo}\SetKwFunction{proc}{proc}
	    \SetKwProg{myfunc}{Function}{}{}
	    \myfunc{$\mathsf{SqORAM\_read}(b)$}{
		$root \gets $ Get root node of ATM from memory\;
		$ATM.path \gets $ Get root to leaf path containing entry for $b$\;
		\While{not at leaf}{
		    $child\_num = ATM.path.nextNode.id$\;
		    $l\_ctr \gets \mathsf{root.getVal(child\_num)}$\;
		    $child\_level \gets \mathsf{Determine\_Level(g, l\_ctr)}$\;
		    $child \gets$ Get Child node from $child\_level$\;	
		    $root = child$\;
		}
		
		$l\_ctr \gets \mathsf{root.getVal(b)}$\;
		$level \gets \mathsf{Determine\_Level(g, l\_ctr)}$	\;		
		$loc\_in\_level \gets \mathsf{level.bufferMap.query}$\;
		$blk \gets$ Read block from $loc\_in\_level$\;			
		\Return $blk$\;
		}
		\myfunc{$\mathsf{SqORAM\_write}(b,d)$}{
		$\mathsf{WriteQueue.push(b,d)}$\;
		$ATM.path\gets$ Root to leaf path containing entry for $b$\;
		$ATM.node = ATM.path.leafNode$\;
		$\mathsf{ATM.node.update(b,g+1)}$\;
		\While {not at root}{
		    $\mathsf{WriteQueue.push(ATM.node.id, 
			ATM.node.data)}$\;
		    $ATM.node = ATM.path.node.parent$\;
		}
		// Flush contents of Write Queue sequentially to top level write buffer\;
		\If {$WriteQueue.full$}{
		    $\mathsf{merge\_deamortized()}$
		    }
		}
\end{algorithm}

\begin{thm}[1]
	%\label{thm:correctness}
The amortized merge procedure (Algorithm~\ref{merge_level}) ensures that all data blocks from the merge and write buffers of level $i-1$ (for any $i \leq logN$) are merged in ascending order of their logical addresses and written to level $i$, within $2^{i}\cdot \beta$ accesses.
\end{thm}

\BPF

W.l.o.g. consider that at a particular stage of the merge, $x \leq 2^{i-1}\beta$ blocks have been written from the {\em write buffer} of level $i-i$ and $y \leq 2^{i-1}\beta$ blocks have been written from the {\em merge buffer} of level $i-1$ to level $i$. Since, none of the buffers have been read entirely ($ctr_{w}, ctr_{m} \leq 2^{i-1}\beta$), both $q_w$ and $q_m$ will contain only real blocks at this stage. Now, consider that in subsequent $2^{i-1}\beta - y$ accesses, all remaining blocks from the level $i-1$ merge buffer are written to level $i$ --  a similar argument holds if blocks from the write buffer are written to level $i$ instead. In this case, $q_m$ will contain fake blocks for the remaining $2^{i-1}\beta - x$ accesses (Step 13).  

Since fake blocks invariably have logical address $N+1$ (greater than the logical address of any real data block), the next $2^{i-1}\beta - x$ 
writes will be from $q_w$ (Step 23), until both $q_w$ and $q_m$ contain fake blocks. Further, either of the buffers in level $i-1$ can contain at most $2^{i-1}\beta$ real blocks and 
fake blocks can be added to $q_w$ only after all real data blocks 
have been written (Step 7). Thus, the remaining real blocks from the write buffer will be necessarily written to level $i$ within the next $2^{i-1}\beta - x$ accesses.
\EPF

\begin{thm}[4]
	The number of disk seeks performed by {\sys} for level reshuffles across $\log{N}$ levels, where $N$ is the number of blocks in the ORAM, amortized over the number of writes is $\frac{4\log{N}}{ \beta}$.
\end{thm}

\begin{proof}
	Consider the process of merging level $i-1$ buffers and writing to level $i$ as per Algorithm \ref{merge_level}. 
	For the merge at least $M > 2 \times \beta$ blocks of memory is required -- to store two buckets entirely, and compare them in linear time. 
	
	First, blocks are read {\em sequentially} from the merge buffer and the write buffer of level $i-1$ into the in-memory queues, $q_m$ and $q_w$  respectively, until the queues are full. Observe that filling up the queues requires only two disk seeks overall -- to place the head at the starting location of the write buffer and then the merge buffer. After the queues are full, $\beta$ blocks are written to the write buffer of level $i$ {\em sequentially}, which requires {\em one} disk seek. Finally, a leaf node is added to the corresponding buffer B-tree map with entries for block that are written to the bucket, and the parent node(s) of the leaf node are updated in memory. If required, the parent node(s) are flushed to the disk and written {\em sequentially} after the leaf node. This requires 1 seek in total.

	Overall, writing a bucket containing $\beta$ blocks to the write buffer requires 4 seeks in total. Since, the write buffer of level $i$ contains $2^{i}$ buckets, the total number of seeks to complete filling up the write buffer is $4 \times 2^{i}$. Also, the write buffer in level $i$  gets filled up after $2^{i}$ writes. Thus, the number seeks for all level reconstructions amortized over the number of writes is:
	
	\begin{center}
		$s = \sum\limits_{i=0}^{\log_{k}{N/B}} \frac{4 \times 2^{i}}{2^{i} \beta} = \frac{4 \log{N}}{\beta} $
	\end{center}
	
\end{proof}

\begin{thm}[5]
	The de-amortized merge protocol (Algorithm~\ref{merge_level_deamortized}) ensures that by the time the write buffer of level $i < \log{N}$ is full, all real data blocks from the merge buffer of level $i$ have been written to level $i+1$.
\end{thm}

\begin{proofsketch}
Every successive execution of Algorithm \ref{merge_level_deamortized} writes $\beta$ blocks to the write buffer of level $i$ {\em sequentially}. At the same time, $\beta$ blocks are written from the merge buffer of level $i$ to the write buffer in level $i+1$. Note that within exactly $2^{i+1}$ successive executions of Algorithm \ref{merge_level_deamortized}, the write buffer of level $i$ will be full. But within the same time, $2^{i+1}\cdot \beta$ blocks will have been written from the merge buffer of level to level $i+1$ (Line 18 - 29).
From Theorem \ref{thm:amort_correctness}, all real blocks from the merge buffer of level $i$ will necessarily be written to level $i+1$ within $2^{i+1}\cdot \beta$ writes. Thus, when the buffers are switched after $2^{i+1}\cdot \beta$ writes, all valid content from the merge buffer of level $i$ will be in level $i+1$.
\end{proofsketch}

\begin{thm}[6]
	The de-amortized merge protocol (Algorithm~\ref{merge_level_deamortized}) ensures that during the merge all writes to level $i$ are 
	uncorrelated and indistinguishable, independent of the logical block addresses.
\end{thm}

\begin{proofsketch} 
	It is straightforward to see that while executing Algorithm \ref{merge_level_deamortized} (in Appendix), the only steps visible to the adversary are Step 29, 32 and 41. 
	
	Step 29 and 41 combined perform a write to the {\em next} block in sequence to the write buffer in every level of the ORAM, irrespective of the logical block address and content. Also,  invariably the process is continued until $\beta$ blocks have been written {\em sequentially} to the write buffer of every level (Step 18). In case, there are less than $\beta$ real data blocks that can be written {\em fake} blocks are written instead. Semantic security ensures that fake blocks are indistinguishable from real data blocks.
\end{proofsketch}

\begin{thm}[7]
	The deamortized {\sys} construction provides write access privacy (Definition \ref{def:privacy}).
\end{thm}

\begin{proofsketch}
The de-amortized construction makes two changes to the amortized
construction: the level reshuffles are performed using the de-amortized merge protocol (Algorithm \ref{merge_level_deamortized}), and the reads are optimized by using the ATM to precisely locate blocks.

Consider two equal-length write access patterns $\vec{A} = {w_1, w_2, \ldots w_i}$ and $\vec{B} = {x_1, x_2, \ldots 	x_i}$. First, note that in the de-amortized construction (similar to the amortized construction) when either of $\vec{A}$ or $\vec{B}$ is executed, the $i$ blocks are added to the write queue irrespective the logical addresses.  Once the write queue is full, its contents are written the top level write buffer, encrypted with semantic security. 

Further, the de-amortized level reshuffling protocol ensures the same security guarantees as the amortized protocol -- while reshuffling level $j < \log N$,  the writes to the disk are uncorrelated to each other and independent of the block addresses (Theorem \ref{thm:deamort_security}). 

Therefore, by simply observing the writes to the top level and the writes due to the level reshuffles, an  adversary can only do negligibly better than purely guessing whether $A$ or $B$ was executed.
  
Second, since reads are not observable, using the ATM during reads to
reduce the number of levels searched for a block does not provide any
additional information to the adversary. Finally, the ATM blocks are
written to the write queue indistinguishably from data blocks due to
semantic security of the encryption.
\end{proofsketch}

\end{document}